%% file: manuscript.tex
\documentclass[sigconf]{acmart}
\copyrightyear{2025}
\acmYear{2025}
\setcopyright{cc}
\setcctype{by}
\acmConference[CHI '25]{CHI Conference on Human Factors in Computing Systems}{April 26-May 1, 2025}{Yokohama, Japan}
\acmBooktitle{CHI Conference on Human Factors in Computing Systems (CHI '25), April 26-May 1, 2025, Yokohama, Japan}
\acmDOI{10.1145/3706598.3713545}
\acmISBN{979-8-4007-1394-1/25/04}

\AtBeginDocument{%
  }

\usepackage{enumitem}
\usepackage{tabularx}

\begin{document}

\title[Prevalence and Impacts of IBSA]{Prevalence and Impacts of Image-Based Sexual Abuse Victimization: A Multinational Study}

\author{Rebecca Umbach}
\email{rumbach@google.com}
\orcid{0000-0002-4463-1096}
\authornotemark[1]
\affiliation{%
  \institution{Google}
  \country{USA}
}

\author{Nicola Henry}
\orcid{0000-0003-2241-7985}
\email{nicola.henry@rmit.edu.au}
\affiliation{%
  \institution{RMIT University}
  \city{Melbourne}
  \country{Australia}
}

\author{Gemma Beard}
\orcid{0009-0006-4197-1956}
\email{gemma.beard@rmit.edu.au}
\affiliation{%
  \institution{RMIT University}
  \city{Melbourne}
  \country{Australia}
}

\renewcommand{\shortauthors}{Umbach et al.}

\begin{abstract}
Image-based sexual abuse (IBSA) refers to the nonconsensual creating, taking, or sharing of intimate images, including threats to share intimate images. Despite the significant harms of IBSA, there is limited data on its prevalence and how it affects different identity or demographic groups. This study examines prevalence of, impacts from, and responses to IBSA via a survey with over 16,000 adults in 10 countries. More than 1 in 5 (22.6\%) respondents reported an experience of IBSA. Victimization rates were higher among LGBTQ+ and younger respondents. Although victimized at similar rates, women reported greater harms and negative impacts from IBSA than men. 
Nearly a third (30.9\%) of victim-survivors did not report or disclose their experience to anyone.  We provide large-scale, granular, baseline data on prevalence in a diverse set of countries to aid in the development of effective interventions that address the experiences and intersectional identities of victim-survivors. 
\end{abstract}





\maketitle

\section{Introduction}
Image-based sexual abuse (IBSA), also known as ``nonconsensual pornography'' or ``nonconsensual intimate images'' (NCII), refers to creating, taking, or sharing intimate images (photos or videos) of someone without their consent. It also includes threatening to share intimate images (``sextortion''), pressuring, coercing, or threatening someone into sharing their intimate images (``sexting coercion''), or sending unwanted intimate images (``cyberflashing''). Since the invention of the first photographic camera in the 1800s, advances in technology have long been used for creating or sharing sexual imagery \citep{coopersmith1998pornography} and for sex work more broadly \citep{barwulor2021disadvantaged, hamilton2022risk}. The internet and the advancements of technological devices have radically democratized digital image creation and distribution, enabling easy sharing through messaging and websites. While computers facilitate human creativity, expression, and exploration in consensual intimate image sharing, unfortunately they also open avenues for misuse and abuse. Social media platforms act as public bulletin boards, messaging, and social network graphs all at once. Online communities provide spaces and opportunities for perpetrators to interact and pursue (dysfunctional) social connections with online peers through the nonconsensual sharing of intimate images \cite{henry2019image}. The integral role of technology in the perpetration of IBSA is reflected in its widespread designation as a form of technology-facilitated gender-based violence (or digital violence) by civil society \citep{tfgbv}, and scholars in criminology \citep{ henry2018technology} and computer science \citep{qiwei2024sociotechnical}. 

Previous research has documented significant harms and impacts associated with IBSA victimization, including anxiety, depression, suicidal ideation, 
reputational and financial damage, as well as other psychological, physical, and social harms \citep{aborisade2022image, bates2017revenge, schmidt2023mental, dardis2022nonconsensual, champion2022examining,campbell2022social, mcglynn2021s}.
Existing research also indicates that IBSA affects a substantial proportion of the population, however there are several methodological and definitional challenges that hinder comparisons between different studies and limit our ability to draw firm conclusions about prevalence. Some studies have broadly investigated the prevalence of IBSA alongside sexting or other technology-facilitated behaviors \citep{powell2019technology, gamez2015prevalence, marganski2018intimate, gasso2020sexting, dunn2023supporting}. Conversely, several studies have solely focused on IBSA, with some exploring a single form of IBSA, such as the nonconsensual distribution of intimate images \citep{brighi2023prevalence, ruvalcaba2020nonconsensual, walker2021nonconsensual, branch2017revenge, esafety2017} or sexting coercion \citep{noorishad2022investigating, ross2019sexting, drouin2015sexting}, and others a range of different IBSA behaviors within a single survey \citep{henry2020image, henry2019image, reed2016snooping, sparks2023image, murcca2023prevalence, marcum2022role, dardis2022nonconsensual}. As a result, the broader studies have not comprehensively examined IBSA, while narrower studies have only examined one form of IBSA.
From a sampling perspective, most studies have measured prevalence within convenience samples, and with one exception \citep{dunn2023supporting}, the prevalence studies on adult victimization have been primarily conducted in four countries, leaving us with a data void in the rest of the world. This limits understandings of IBSA and impairs our ability to compare experiences across demographic and identity characteristics. 

The body of work on prevalence highlights several key challenges in addressing this online harm, namely the evolving definition of IBSA behaviors, and a lack of quantitative, representative research in countries outside of Australia, New Zealand, the UK, and the United States \citep{linxen2021weird}.

The relationship between victim and perpetrator has also been the subject of relatively little investigation to date, although existing findings consistently find IBSA perpetrators are most commonly current or former intimate partners \citep{wei2024understandinghelpseekinghelpgivingsocial, powell2019image, ruvalcaba2020nonconsensual, karasavva2022personality, dardis2022nonconsensual, esafety2017}. Understanding context is critical for developing interventions to prevent and reduce IBSA. This includes a better understanding of the role that demographic and identity characteristics play in the dynamics of the experience (e.g., gender of the victim-survivor; relationships between perpetrator and victim-survivor), and the consequences. For instance, educational interventions focused on abstinence may fail, especially given the ubiquity and normalization of sexting in romantic relationships \citep{Power2022}. 

Finally, another literature gap pertains to large-scale data concerning how people seek help or report their IBSA experience. A common finding across studies is the reluctance to report or disclose experiences of IBSA \citep{ruvalcaba2020nonconsensual, brighi2023prevalence, henry2024wasn, wei2024understandinghelpseekinghelpgivingsocial, esafety2017}. Those who disclose may only do so anonymously on social media or to people close to them, rather than reporting to the police or going to online safety or victim support organizations \citep{patchin2020sextortion, wei2024understandinghelpseekinghelpgivingsocial}. Given the limited data on help-seeking and reporting, expanding this knowledge is critical for the development of effective prevention and support interventions and tools. 

We investigated the prevalence and impacts of different types of IBSA victimization, as well as victim-survivors' experiences of help-seeking and reporting. To do so, and to address the gaps and limitations in the literature identified above, \color{black} our research was guided by the following research questions: 

\begin{itemize}
\item\textbf{RQ1:}~\textit{What is the prevalence of IBSA victimization by country?}
\item\textbf{RQ2:}~\textit
{What is the relationship between IBSA victimization and demographic characteristics, including age, gender, and sexual minoritization?}
\item\textbf{RQ3:}~\textit
   {What is the nature of the relationships between victim-survivors and perpetrators of IBSA?}
\item\textbf{RQ4:}~\textit
{What are the impacts on victim-survivors of IBSA?} \item\textbf{RQ5:}~\textit
{What are the help-seeking and formal reporting experiences of IBSA victim-survivors?}
\end{itemize}

Using the largest and most diverse sample to date, as well as representative sampling and appropriate weighting techniques, this study investigates IBSA prevalence for 10 countries, eight of which have no existing prevalence data to date. An online survey was administered to over 16,000 respondents, sampled representatively with regard to age, gender, and region. Respondents were asked about their own experiences of IBSA perpetration and victimization. In this paper, we report on the latter, looking at experiences where respondents reported that someone nonconsensually created, filmed, photographed, shared, or threatened to share, nude or sexual images of them. In addition to our quantitative analysis of the survey responses, we also used qualitative analysis for open-ended questions to add context to our findings, particularly on impacts, help-seeking, and reporting. Our analysis shows that almost a quarter of respondents (22.6\%) report at least one victimization experience. While overall rates of IBSA are identical for men and women, we found elevated rates of victimization for LGBTQ+ men and women, and young people. The majority of victim-survivors reported the perpetrator was a current or former partner. Perceptions of harms and impacts as a result of victimization differed by gender, with women reporting greater harms than men. Inclusive of platform reporting options, 38.8\% of victim-survivors had reported, 14.3\% has disclosed but not reported, 30.9\% had not reported or told anyone, and the remaining 15.9\% chose not to respond to this question.

Effectively combating online harms requires a clear understanding of their scale and scope. Technology-facilitated gender-based violence is often under-reported or poorly understood. Our finding that almost a quarter of respondents report an experience of IBSA victimization, which represents an increase from previously reported rates, highlights the critical need for intervention, both with regard to primary prevention and recovery.  

We used an inclusive IBSA definition, including behaviors that have rarely been investigated, and report both aggregated and disaggregated prevalence numbers. 
This facilitates comparisons to previous or future studies, which may focus on a single subtype of IBSA. 
 We shed light on how demographic and identity risk characteristics influence the likelihood of IBSA experience, the harms or impacts felt by victim-survivors, and importantly, help-seeking and reporting behaviors - all of which are under-researched in the current literature. By offering insights that are applicable across different cultural and legal context, we advance scholarship on IBSA, and provide actionable insights for legislators, researchers, and technology companies for prevention and mitigation efforts. Overall, our study makes important contributions to the HCI literature by firstly investigating the prevalence, nature, and impacts of IBSA, an important and growing HCI problem (e.g., individuals using digital technologies which cause harm), and secondly making recommendations for how HCI interventions can be harnessed to both prevent and better respond to IBSA.

\section{Related Work}

In this section, we summarize the key findings from the research literature on prevalence, impacts, and help-seeking regarding the following forms of IBSA: the nonconsensual sharing of intimate images; the nonconsensual taking of intimate images; the nonconsensual creation of intimate images; and making threats to share intimate images. We exclude sexting coercion and cyberflashing as these forms of IBSA were not explored in our survey.

\subsection{Prevalence and Risk Factors}

First, regarding the nonconsensual \textit{sharing} of intimate images, as detailed in a review by \citet{paradiso2023image}, there are mixed results from existing studies on both baseline prevalence, and whether rates of IBSA victimization differ by gender. Rates varied between $1.1\%$ \citep{gamez2015prevalence} and $28.5\%$ \citep{karasavva2022personality}, with some studies showing either roughly comparable rates between men and women in their samples \citep{gasso2020sexting}, and others showing higher rates for men \citep{henry2019image, powell2019technology, henry2020image, dunn2023supporting} or women \citep{walker2021nonconsensual, dardis2022nonconsensual, ruvalcaba2020nonconsensual, esafety2017}. 
However, several of these studies used non-representative sampling methods, such as convenience sampling of university students. Few studies have investigated demographic risk factors for nonconsensual sharing victimization beyond gender, including sexuality, age, and/or race. Moreover, while some studies examining online abuse more generally report on those risk demographics, they do not always do so in relation to IBSA specifically or specific IBSA behaviors. Of the studies that have reported on demographic risk factors for nonconsensual sharing victimization, higher rates of victimization are reported by lesbian, gay, and bisexual respondents, relative to heterosexual respondents \citep{ruvalcaba2020nonconsensual,brighi2023prevalence, henry2019image, gamez2015prevalence, dunn2023supporting, esafety2017}. The research also shows that ``emerging adults'' (those aged between $18-29$) are more likely compared to older adults to have experienced IBSA \citep{ruvalcaba2020nonconsensual, henry2020image, gamez2015prevalence, esafety2017}, as were Indigenous, Black, and other minoritized races or ethnicities \citep{ruvalcaba2020nonconsensual, henry2020image, esafety2017}.

Second, only three studies have explored the nonconsensual \textit{taking (photographing or filming)} of intimate images, with prevalence ranging from $10.7\%$ to $33.2\%$ \citep{powell2019image, henry2019image, henry2020image}. All three studies found similar results to those of nonconsensual sharing, with men, sexual minorities, emerging adults aged 18-29, and racial or ethnic minorities disproportionately affected. This form of IBSA is one of the most under-researched in the existing literature.

Third, in relation to the nonconsensual \textit{creation} of intimate images, including creating AI-generated images or digitally altering images (including ``deepfake'' images), there is also limited empirical research to date. \citet{flynn2022deepfakes} found that men, 
younger people, those with a disability, and sexual minorities were more likely to report that digitally altered intimate images of them were created without their consent. 
In a study by \citet{umbach2024non}, $1.2\%$ of respondents said that someone had created deepfake nude or sexual images of them without their consent, with men reporting higher rates than women.

Finally, little prevalence data exists regarding \textit{threats to share} intimate images (or ``sextortion''). Across studies, rates of sextortion against adults ranged between $6.8\%$ to $18.7\%$ \citep{brighi2023prevalence, henry2019image, henry2020image, powell2019image, henry2024sextortion}. Findings were similar to that of nonconsensual creation --- that is, disproportionate rates of victimization in men \citep{powell2019image, henry2019image, henry2020image,henry2024sextortion, umbach2024non}, sexual minorities, young people, and 
racial or ethnic minorities \citep{henry2020image, eaton2023relationship, henry2024sextortion}.
 
Empirical findings in the scholarly literature on IBSA present mixed results. Much of the existing research focuses on the actual sharing of intimate images, as opposed to other behaviors, such as surreptitious filming, creating ``deepfakes,'' or threats to share intimate images. Emergent research indicates that certain subtypes of IBSA show higher victimization rates in specific populations, underscoring the need for more focused investigations into both general prevalence, as well as probing the effects of identity characteristics. This motivates our first two research questions:
\begin{itemize}
\item\textbf{RQ1:}~\textit{What is the prevalence of IBSA victimization by country?}
\item\textbf{RQ2:}~\textit
{What is the relationship between IBSA victimization and demographic characteristics, including age, gender, and sexual minoritization?}
\end{itemize}

\subsection{Relationships}
Of the few studies that have examined the relationship between the victim and perpetrator, all found that men disproportionately perpetrated IBSA, and our findings that perpetrators were typically current or former romantic partners at the time of the abuse \citep{branch2017revenge, ruvalcaba2020nonconsensual, henry2020image, dardis2022nonconsensual, umbach2024non, henry2024sextortion, esafety2017}. One study \citep{wei2024understandinghelpseekinghelpgivingsocial}, examining Reddit data related to IBSA help-seeking, observed the same patterns around current and past intimate partners for nonconsensual distribution of images and non-financially motivated sextortion. Strangers were more likely to engage in financial sextortion and cyberflashing. Understanding the nature of the relationships between perpetrators and victims is crucial, informing interventions for primary prevention (e.g., what threat models are most salient before, during, and after an experience?), and contextualizing what actions victim-survivors might take, or want to take, after an experience. Therefore, our third research question asks: 
\begin{itemize}
\item\textbf{RQ3:}~\textit
  {What is the nature of the relationships between victim-survivors and perpetrators of IBSA?}
\end{itemize}

\subsection{Harms and Impacts}
Few quantitative studies have investigated the harms and impacts of IBSA, including negative or positive moods, emotions, feelings, responses, or other outcomes. Studies report a range of harmful impacts. For instance, \citet{champion2022examining} found that $71.0\%$ ($n=22$) of survey respondents who reported IBSA as their most impactful technology-facilitated sexual violence experience indicated that it caused high levels of stress and anxiety, with some respondents reporting depression, problematic alcohol use, and/or suicidal ideation. In addition to psychopathological effects, other studies investigated feelings, concerns, emotions, or other outcomes. \citet{dardis2022nonconsensual} noted that victims of IBSA reported a range of negative emotions, in particular, betrayal, anger, and concern. In \citet{henry2020image}'s study, of the $957$ victim-survivor respondents, $61.1\%$ ($n=585$) reported negative feelings as a result of their most significant experience of IBSA (e.g., feeling embarrassed, humiliated, ashamed, depressed, angry, fearful for safety, or concerned about reputation), whereas 38.9\% ($n=372$) reported neutral or positive feelings or outcomes (e.g., they felt ok, flattered, or found it funny). The study also asked respondents about the physical and psychological impacts of their most significant experience of IBSA, finding that more than half reported harms to health and relationships, and over a third experienced harassment. 

More research is needed to understand the physical, psychological, and social consequences of IBSA. In particular, more research is needed to understand how the harms and impacts might be different depending on age, gender, race/ethnicity, sexuality, ability, and other factors. In the study by \citet{ruvalcaba2020nonconsensual}, they found significant differences in psychological wellbeing between women victims and women non-victims, but no differences were found between men victims and men non-victims. In \citet{gasso2020sexting}, they found that men respondents who were victims of nonconsensual dissemination were over five times more likely to present global psychopathology compared to their non-victim counterparts, with no effects on depression and anxiety. Women were over two times more likely to show depression, anxiety, \textit{and} global psychopathology compared to non-victims. In \citet{henry2020image}, they found that women were significantly more likely than men to report negative or detrimental feelings or outcomes, including negative health and relationship impacts, as well as reputational and safety concerns. Similarly, in \citet{dunn2023supporting}, women were more likely than men to report nonconsensual sharing to be extremely harmful.

In addition to gender, one study also looked at age, race/ethnicity, and sexuality as moderators of experience. In their multi-country study, \citet{henry2020image} found that White, European, and Pākehā women were more likely than women from minoritized racial/ethnic communities to report negative feelings or reputational or safety concerns, although the women from minoritized groups were more likely to report relational harms, as were LGB+ women. The study did not find significant differences between men based on race/ethnicity, or between different age or sexuality groups. Finally, men, those from racial/ethnic minorities, LGB+ women, and those aged $16-19$ were more likely than their counterparts to report experiencing harassment as a result of their most significant experience of IBSA.

Qualitative research, involving interviews with victim-survivors of IBSA, has also investigated victim-survivor impacts. Across the different studies, researchers have found that there are often significant mental health impacts associated with IBSA, including post-traumatic stress, suicidal ideation, depression, and anxiety, as well as social impacts such as loss of control, social isolation, erosion in self-esteem, self-blame, stigmatization, lack of trust, hyper-vigilance, fear of re-victimization, and harmful coping mechanisms \citep{bates2017revenge, aborisade2022image, campbell2022social, mcglynn2021s, champion2022examining}. 

Overall, existing research suggests that identity characteristics may influence victimization experiences, wherein the harms may be compounded by intersecting forms of marginalization. Many of these studies focused specifically on psychopathology, while others pointed to wide-ranging impacts that touch many aspects of a victim-survivors' life. A more fulsome picture of impacts and harms can inform programs and services for victim-survivors, and help with prioritization efforts among law-makers. Our third research question is as follows:
\begin{itemize}
\item\textbf{RQ4:}~\textit
  {What are the impacts on victim-survivors of IBSA?}
\end{itemize}

\color{black}
\subsection{Help-Seeking and Reporting}
Finally, several qualitative and quantitative studies have investigated victim-survivor reporting and help-seeking experiences, although this has received comparatively less attention within the research literature. \citet{ruvalcaba2020nonconsensual} conducted exploratory analyses on victims' help-seeking behaviors, finding that most respondents who experienced IBSA did not turn to anyone for help ($72.95\%$). Common reasons for not reporting included embarrassment, fear, ``it didn't bother me,'' or ``I didn't have time.'' In the study by \citet{brighi2023prevalence}, $43.9\%$ of participants had not spoken to anyone about what had happened, which was higher for sexual minority victims. The most common reason for non-disclosure was embarrassment, followed by feeling like talking about it would not have helped. Those who did disclose most commonly did so to close friends. In \citet{henry2020image}, only one third of IBSA victim-survivors were confident they knew where or how to access support. This varied by country, with UK respondents being less confident compared to those in Australia and New Zealand.
It is imperative to understand the barriers to and motivations for reporting. This can shape the design of resources such as chatbots \citep{maeng2022designing, tan2024scoping}, forms to request removal from sites and search results \citep{de2021reporting}, training for help-givers \citep{zou2021role}, and educational materials. It can also help to identify how existing resources are used, as well as inform punishment for perpetrators. This leads us to our final research question:

\begin{itemize}
\item\textbf{RQ5:}~\textit
  {What are the help-seeking and formal reporting experiences of IBSA victim-survivors?}
\end{itemize}

Overall, there are several challenges in drawing comparisons on prevalence, impacts, reporting, and help-seeking behaviors across the scholarly literature on IBSA victimization. First, study sample size and approaches to recruitment and sampling were highly variable. Second, studies employed different phrasing and terminology. For instance, some studies referred to ``posting,'' ``sharing,'' ``sending,'' or ``distributing'' which may have different meanings. Some studies only asked about images being shared online, while others focused on email or text. There was also divergence in the use of terms such as ``sexual,'' ``nude,'' ``sexually explicit,'' and ``intimate,'' which can mean different things. Moreover, some studies asked more broadly about intimate or sexual images as well as other content. Finally, not all quantitative studies measured prevalence, and not all quantitative or qualitative studies investigated impacts, reporting, and/or help-seeking behaviors. 
\section{Methods}
In mid-2023, we conducted a multi-country online survey on IBSA, defined as the nonconsensual taking, creating, or sharing of intimate images, including threats to share intimate images. In this paper, we report on findings in relation to victimization, in order to contribute further knowledge about adult experiences of different types of IBSA.

\subsection{Ethical Considerations}
 The study procedure and survey instrument were approved by the Human Research Ethics Committee at
 RMIT University. At the beginning of the survey, respondents were given information about the topic and could opt out. All questions involving self-reporting of victimization and/or perpetration were voluntary and respondents could choose to answer ``prefer not to say'' for any of these questions. At the end of the survey, respondents were given information about resources, including reporting to the police or contacting a support service. Respondents were compensated for their time in their local currency (decided and paid by YouGov, the panel provider).
\subsection{Study Design}
The online survey was administered in 10 countries. The countries included: Australia, Belgium, Denmark, France, Mexico, the Netherlands, Poland, South Korea, Spain, and the United States. We selected countries based on a mixture of considerations. First, some lack IBSA legislation (Poland, the Netherlands, Denmark) while others have existing or pending legislation on IBSA (Australia, Belgium, France, Mexico, United States, Spain, South Korea). Second, we chose these countries because of their geographical diversity. And third, two countries had prior research findings on IBSA  (Australia, United States), while the remainder did not. This enabled comparisons to be made for some countries as well as proffering new findings in other countries where no previous data exists. In the interest of participant protection, we also refrained from fielding the survey in countries where answering such a detailed survey may be dangerous or perceived as such for respondents, or where there were a lack of local resources where we could direct respondents. 

YouGov, a market research and data analytics firm, was used to administer the survey to a minimum of 1,600 adults (18+) drawn from their panels in each of the countries for a total respondent count of 16,693. YouGov has global coverage with large panels \citep{reutersyougov}, which reduces the likelihood of needing extensive within-country weighting. Screener questions were used to achieve representative sampling based on quotas. Samples were designed to be representative by age, gender, and region according to the latest official population estimates of each country. Upon qualifying for the survey, completion rates ranged from 71.0\% (Belgium) to 89.0\% (Australia). The final dataset consisted of respondents who completed the entire survey and passed the YouGov standard quality checks, including bot catchers, speeder detection, and manual checking of open-text questions.

The survey (median completion time 18.2 minutes) built on an earlier survey conducted in Australia by \citet{powell2019technology}. The survey began with a description of the topic. Individuals then answered a set of screener questions on country of residence, age, and gender, followed by a consent form. Several general behavioral questions situated the respondent before the survey moved into personal experiences about victimization (including impacts and help-seeking), followed by ``bystander'' questions, which asked about seeking out or receiving nonconsensual imagery. The next two sections were about perpetration and IBSA attitudes, respectively. The survey concluded with several more demographic questions. The survey contained three open-ended questions, which were optional response. In addition, many of the close-ended questions had ``other'' options, which invited (but not did not require) respondents to elaborate. \color{black} In this paper, we focus on the subset of questions that asked specifically about victimization. 
The survey vendor translated the survey into the official language of each country, which were then reviewed by at least one native speaker colleague of the research team for accuracy and cultural sensitivity. The goal was to maintain consistency across translations while accommodating cultural nuances related to sexual topics. 

The mean participant age was 46.0 years (\(sd=16.7\)). Women accounted for 50.9\% of the respondents, men for 47.6\%, and together ``prefer to self describe,'' ``prefer not to say,'' and ``non-binary'' accounted for the remaining 1.4\%. Detailed demographic breakdowns by country are available in Appendix Table A1. When we focus on gender, we do so primarily with respondents who identified as women or men, excluding the other respondents for two reasons: first, our total sample (unweighted) of non-binary respondents was 93, and second, we were limited by our census weighting, which are often restricted to binary genders. When focusing on sexual minoritization as a risk factor, we divided the respondents into an LGBTQ+ group ($n=1,675$) and a non-LGBTQ+ ($n=14,175$) group (excluding a small subset of respondents who could not easily be categorized into either group, $n = 147$). To create the dichotomous variable, we assigned anyone who self-identified as non-binary, gay, lesbian, bisexual, and/or transgender into the former category. We also translated the 327 ``Prefer to self describe'' open-ends via the Google Translate API, and manually coded those respondents. Those who self-described as straight or heterosexual were placed into the non-LGBTQ+ group. The majority (e.g., respondents who indicated pansexuality, demisexuality, or queer) were assigned into the LGBTQ+ group. Following coding, less than 1\% (\(n = 147\)) of respondents remained uncategorized (including those self-identifying as ``normal'').

Respondents who had at least one experience of victimization were asked to report on their \textit{most significant experience} in the resulting follow-up questions.   
A question about which actions, if any, the respondent had taken in response to their experience was used to divide victim-survivor respondents into four distinct groups to answer RQ5: 
(a) ``reporters'': those who reported to the police, contacted an online safety agency, or contacted a website or search engine to request that the content be taken down, and who may or may not have disclosed additionally to friends or family; (b) ``only disclosers'': those who told friends or family or disclosed to someone, but had not formally reported; (c) ``non-disclosers'': those who had told no one nor made any reports; and (d) those who chose not to answer the question (e.g., selected ``prefer not to say'').

\subsection{Questionnaire Design}
The victimization section of the survey asked respondents whether, since the age of 18, someone had: 
\begin{enumerate}[label=\alph*)]
  \item photographed or filmed a nude or sexual image (photo or video) of them without their permission;
     \item created an image of them without their permission 
     where their face and/or body were digitally altered to represent them in a nude or sexual way; 
    \item stolen a nude or sexual image
    of them from their computer, device, or online account;
    \item threatened to post, send, or show a nude or sexual image 
    of them;
    \item posted, sent, or showed a nude or sexual image
    of them without their permission.
\end{enumerate}

Respondents who indicated at least one IBSA experience then answered questions about their \textit{most significant and impactful experience}, detailing the specific acts involved, their emotional response, any resulting harms or impacts, and actions taken. 
Both ``prefer not to say'' and open-ended ``other'' options were available throughout. 

\subsection{Data Analyses}
In some countries, samples were not fully representative. Accordingly, when presenting country-level findings, we use post-stratification weights for age, gender, and location  derived from the most recent census data. When we aggregate across countries or present data related to gender or sexual orientation, we also apply population size weights, rounded to the nearest 1,000. The weighted statistics are presented throughout.

In calculating rates of prevalence, we only counted individuals who indicated at least one victimization experience. Individuals who answered ``prefer not to say'' for all of the victimization questions were not considered victim-survivors. Thus, while the numerator was ``number of people who reported at least one experience,'' the denominator was the total number of survey respondents. Choosing prefer not to answer for all of the victimization questions was relatively rare, ranging from 3.0\% in the United States to 0.9\% in the Netherlands and Australia. 

 All analyses were conducted by the authors. Quantitative data analyses were conducted using \citet{Rstudio}. Due to weighting, we primarily used the survey \citep{lumley}, wCorr \citep{Bailey}, srvyr \citep{ellis}, marginaleffects \citep{arel2023marginal}, jtools \citep{jtools}, and epitools \citep{aragon} packages. When comparing dichhotomized groups (e.g., gender, LGBTQ+/non-LGBTQ+) with dichotomized outcomes, we primarily use and report weighted, adjusted risk ratios calculated from logistic regressions using the svyglm (\citep{lumley}) and avg\_comparisons (\citep{arel2023marginal}) functions \citep{bieler2010estimating}. 

 For the qualitative analysis, we thematically analyzed the open-ended responses to questions in the survey about impacts, help-seeking, and reporting. For instance, respondents were asked how they felt about their experience, what actions they took, and what was the most helpful action. For the analysis we adopted an iterative, inductive approach. First, we translated the responses from the origin language to English via the Google Translate API. Second, the second and third authors read through a sample of responses across the different countries and independently drafted a set of codes, followed by discussion and finalization of the codebook. In the third stage, we independently coded the data using the codebook. Fourth, we then compared coding allocations for quality assurance. Fifth, we compared and contrasted the two coders' spreadsheets, merging them to create a comprehensive master spreadsheet. At this stage, we also identified and removed responses that were deemed inappropriate or irrelevant (e.g., ``I have not experienced any of the above'' or ``not sure''). The final codebook contained 103 codes that focused on impacts and harms, help-seeking and reporting actions, and the perceived helpfulness of respondents' chosen actions. The qualitative part of this study excludes detailed statistics because the qualitative insights are intended to complement the already comprehensive quantitative data obtained in the survey. This approach allowed us to explore nuanced themes and personal experiences that provide depth and context, enhancing our understanding beyond what the quantitative data presents. 
\subsection{\textbf{RQ1: Prevalence of IBSA Victimization by Country}}
\begin{table*}[!htpb]
    \centering
    \input{prevalenceoverall}
    \caption{Prevalence of IBSA victimization by country and by gender. Within-country Bonferroni corrected differences in gender shown in column three, and between-country Bonferroni corrected differences in overall victimization shown in column five via text treatments. \\{*}Note: \textbf{bolded}$<0.05$, \textit{italicized}$<0.01$, \underline{underlined}$<0.001$}
    \label{tbl:prevalence}
\end{table*}
\section{Results}

Nearly 1 in 4 ($22.6\%, 95\%CI=21.5-23.6$) respondents indicated experiencing at least one of the five types of IBSA. A Pearson's $\chi^2$ test with Rao-Scott adjustment confirmed the significant relationship between country and victimization rate ($F=17.53, ndf=9.24, ddf=154268.18, p<0.001$). Rates by country and by gender can be seen in Table \ref{tbl:prevalence}, alongside Bonferroni corrected $p$-value between-country (column 5) and between-gender comparisons (bolding in column 3). As an example, overall rates in Belgium$_b$, and in any other country where a, e, and j are present in column 5, are significantly lower than rates in Australia$_a$, Mexico$_e$, and the United States$j$. Rates were highest in Mexico, and lowest in the European countries. In total, 13.9\% ($95\%CI=13.0-14.8$) of respondents (not shown) indicated at least two different victimization types (e.g., images created, taken, stolen, shared, or threatened).

\begin{figure*}[!htbp]
	\centering
	\includegraphics[scale=.7]{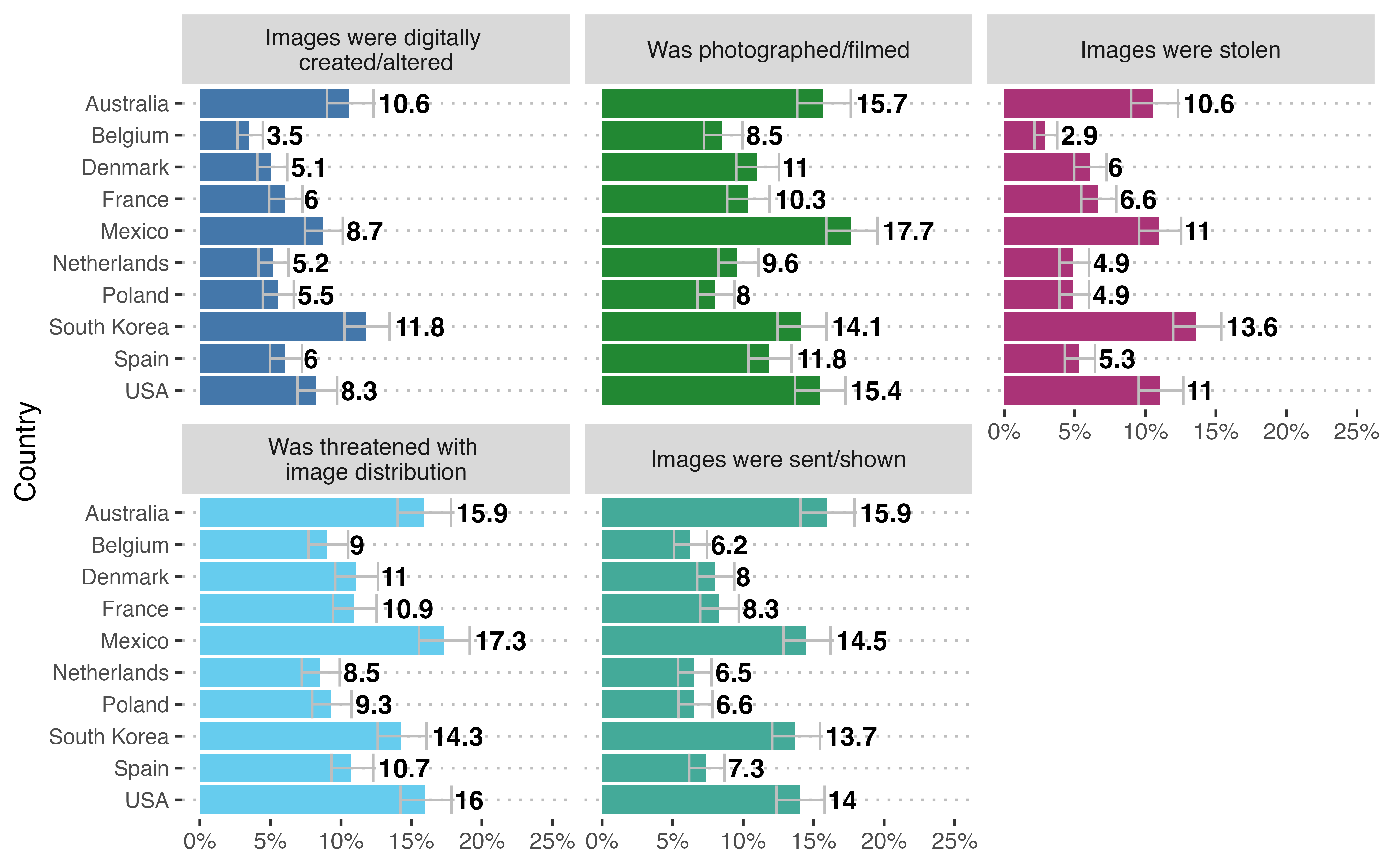}
	  \caption{Faceted bar chart of different types of IBSA, with prevalence by country (95\% CIs in grey)}\label{fig1}
\end{figure*}
Table \ref{tbl2} shows the rates of victimization across the different types of IBSA. These statistics are further broken out by country in Figure \ref{fig1}, with rates ranging from 2.9\% (stolen content in Belgium) to 17.7\% (having content photographed or filmed without consent in Mexico).

\begin{table}[!htbp]
\sf\centering
\caption{Prevalence of subtypes of IBSA across countries
}\label{tbl2}
\begin{tabular}{lrr}
\toprule
\textbf{Type of IBSA} & \textbf{Self-report (95\% CI) }
\\
\midrule
 Created & 8.0\% (7.3-8.7)
 \\
 Photographed/filmed & 14.2\% (13.3-15.1) 
 \\
 Stolen & 9.7\% (8.9-10.5) 
 \\
 Threatened & 14.5\% (13.6-15.4) 
 \\
 Sent/shown & 12.3\% (11.5-13.2) 
 \\
\bottomrule
\end{tabular}\\[10pt]
\end{table}

\subsection{\textbf{RQ2: Relationship between IBSA Victimization and Demographic Characteristics, including Age, Gender, and Sexual Minoritization}}

\subsubsection{Gender}
Aggregated across countries and types of victimization, our study found that men and women were victimized at equivalent rates ($ARR = 1.02, 95\%CI = 0.93-1.12, p=0.72$): 22.7\% ($95\%CI=21.2-24.4$) of men reported at least one type of victimization, while 22.3\% ($95\%CI=21.0 -23.7$) of women reported the same. When broken out by country, as seen in Table \ref{tbl:prevalence}, South Korean women were 1.77 times ($95\%CI = 1.43-2.19, p<0.001$) more likely than South Korean men to report some form of victimization, and Belgian men were 1.39 ($95\%CI = 1.11-1.75, p<0.05$) times more likely than women to report victimization. 
 Broken out by IBSA subtype (Table \ref{tbl3}), women and men were equally likely to report having been photographed or filmed, and having their content sent or shown without their consent. Women were significantly less likely than men to report having content of them digitally-altered or created, being threatened with the dissemination of their content, or having their content stolen via unauthorized access.

\input{genderxsubtypes}
\subsubsection{LGBTQ+ Status}

LGBTQ+ respondents were 1.88 times ($95\%CI =1.69-2.10, p<0.001$) more likely to report IBSA victimization compared to non-LGBTQ+ respondents (38.5\% vs 20.5\%). There were no significant differences between the rate of victimization reported by LGBTQ+ men ($38.5\%$) and LGBTQ+ women ($39.8\%, ARR = 1.03,95\%CI=0.85-1.26, p=0.85$, not shown).

\subsubsection{Age}

Our study found that young people report IBSA victimization at higher rates than older people, as indicated in Table \ref{age}. Visual examination of the five groups show a clear drop off after the age of 34, so the groups were dichotomized into an ``under 35'' and ``over 35'' variable. 
\begin{table}[!htbp]
\sf\centering
\caption{Prevalence of IBSA by age}\label{age}
\begin{tabular}{lr}
\toprule
\textbf{Age group} & \textbf{Prevalence (95\%CI)}\\
\midrule
18-24 & 32.7\% (29.0-36.4) \\
25-34 & 33.0\% (29.8-36.3)\\
35-49 & 19.7\% (17.5-22.0)\\
50-64 & 13.4\% (11.9-15.1) \\
65+& 13.5\% (12.6-14.4) \\
\bottomrule
\end{tabular}\\[10pt]
\end{table}
Approximately a third of respondents under 35 reported experiencing this type of abuse, making them 2.09 times ($95\%CI =1.91-2.29, p<0.001$, not shown) more likely to report victimization as compared to respondents aged 35+. 

\subsection{\textbf{RQ3:The Nature of the Relationships between Victim-Survivors and Perpetrators of IBSA}}
Respondents who indicated at least one victimization experience (unweighted $n=3,189$) were asked to complete follow-up questions about the relationship between themselves and the perpetrators, the harms they experienced, and the actions they took in response, if any. 
For the remainder of the paper, the denominator is the number of respondents who indicated at least one experience of victimization. 

\begin{figure}[!htpb]
	\centering
	\includegraphics[scale=.6]{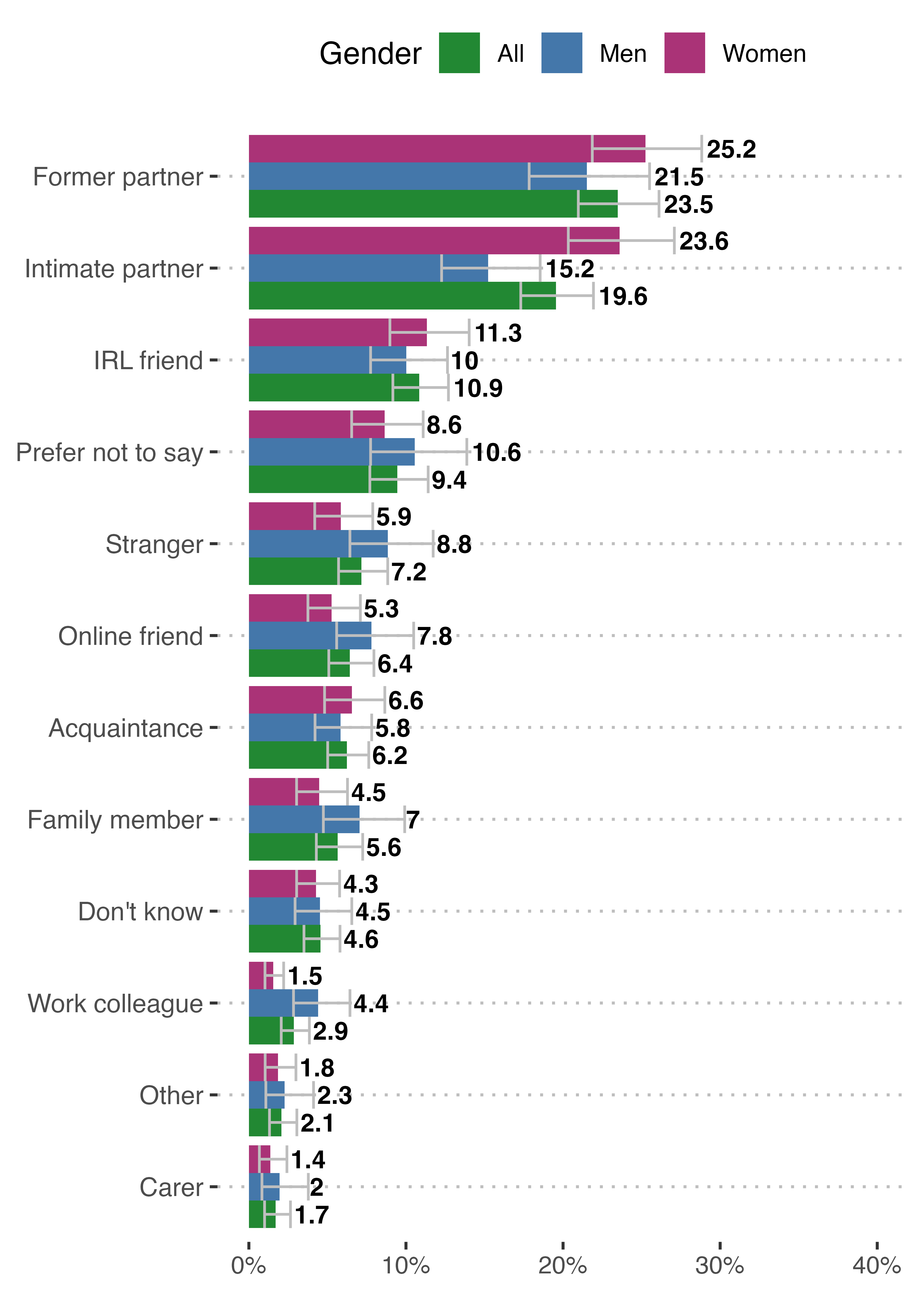}
	 \caption{Best description of relationship to the perpetrator at the time of the incident (95\% CIs in grey).}\label{figrelationships}
\end{figure}

Reported perpetrator gender was disproportionately a man (56.6\%) as compared to a woman (30.5\%). Another 6.9\% were unsure of the gender of the perpetrator, 4.1\% chose not to respond, and the remaining said the perpetrator was non-binary (1.4\%) or something else (0.6\%). Looking at just cases with binary gender perpetrators, the proportion of men perpetrators (65\%) was significantly higher than women perpetrators (35\%, $z=129.48, p<0.001$).
As shown in Figure \ref{figrelationships}, the majority of perpetrators are known to the victim, and almost half are either former and current intimate partners. 
 The only gender differences were that women were more likely to report victimization by a current intimate partner ($ 23.6\%$ vs $15.2\%, ARR=1.55, 95\%CI=1.21-1.99, p<0.001$), and less likely to report victimization by a work colleague ($1.5\%$ vs $4.4\%, ARR=0.35, 95\%CI=0.20-0.61, p<0.001$), as compared to men. There were no significant differences between the LGBTQ+ group and the non-LGBTQ+ group.

\subsection{\textbf{RQ4: The Impacts on Victim-Survivors}}

\begin{figure}[!htpb]
	\centering
	\includegraphics[scale=.6]{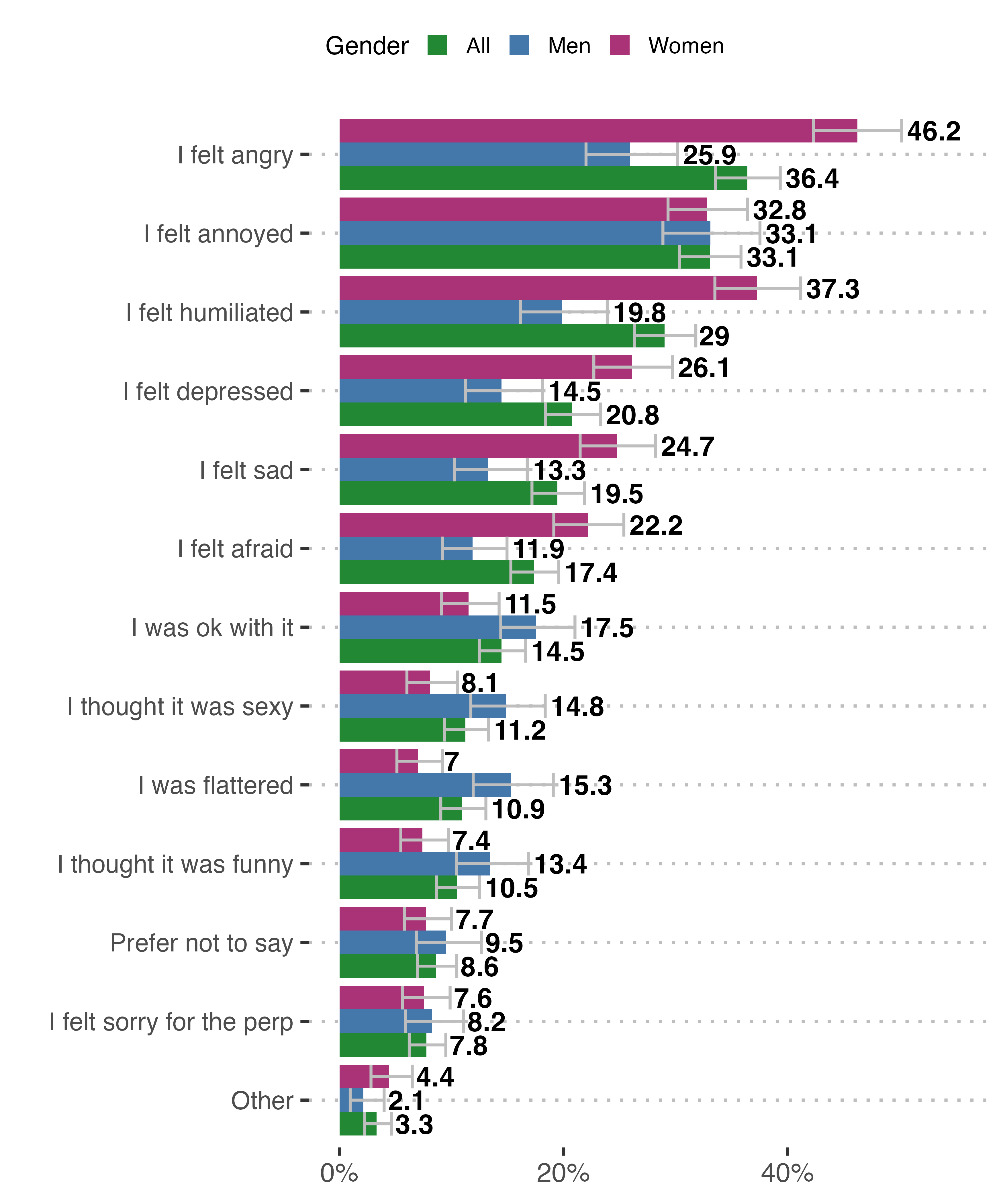}
	  \caption{Feelings experienced as a result of respondents' most significant experiences (95\% CIs in grey).}\label{figfeelings}
\end{figure}

\begin{figure}[!htpb]
	\centering
	\includegraphics[width=.5\textwidth]{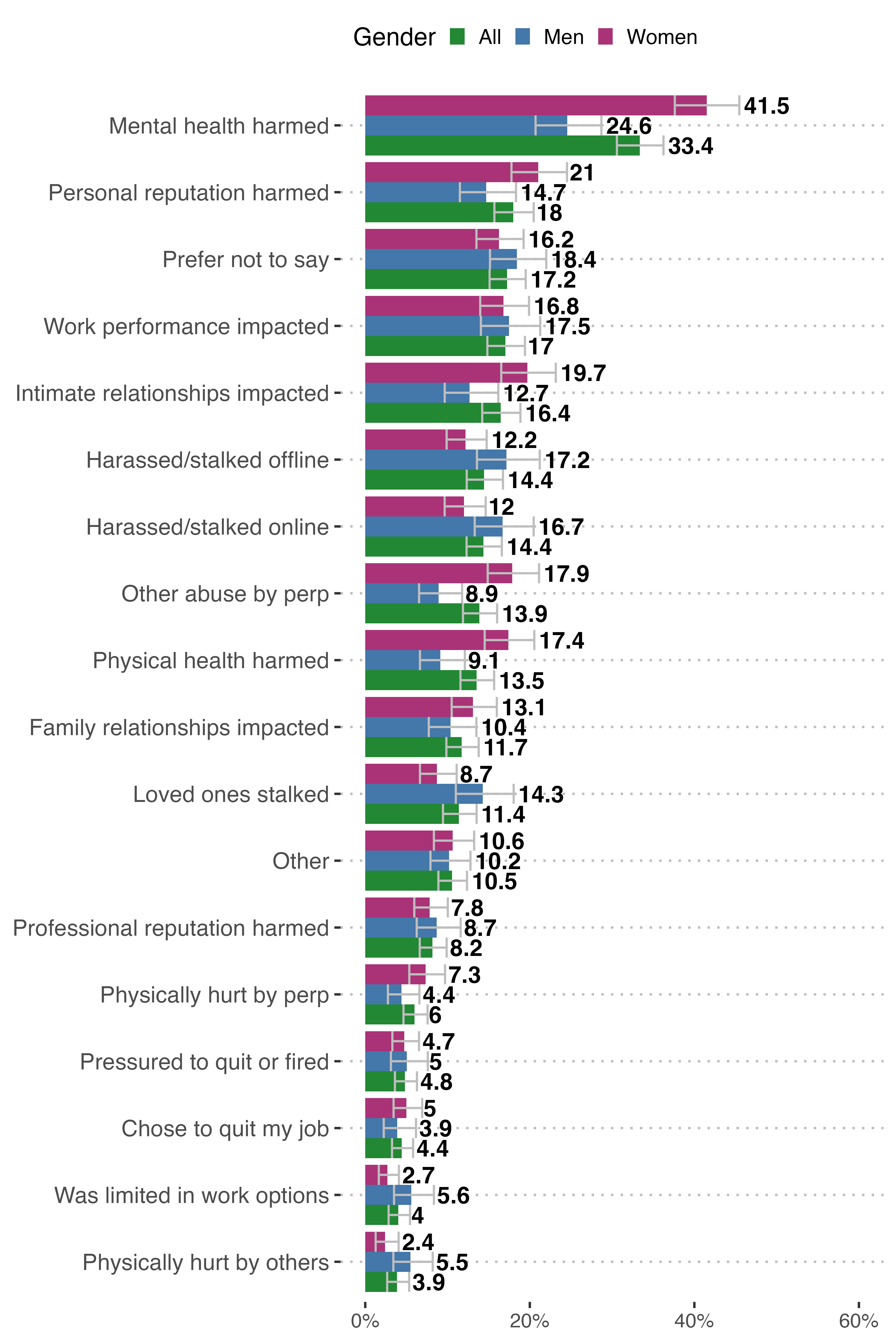}
	  \caption{Harms experienced as a result of respondents' most significant experience
	  (95\% CIs in grey)} \label{fig:harmsexperienced}
\end{figure}

In relation to impacts, the most commonly expressed feelings were annoyance and anger, as seen in Figure \ref{figfeelings}. Women respondents were more likely to report negative feelings such as anger, humiliation, depression, sadness, and fear due to their victimization, while the men were more likely to report being OK with it, or to find it funny, sexy, or flattering. 

We also asked respondents about what harms, if any, they experienced as a result of the abuse, finding that 61.4\% ($95\%CI = 58.8-64.0$) reported at least one harm (excluding ``prefer not to say'' and ``other''). The most commonly reported harm, aggregated across gender, was negative impacts to mental health ($33.4\%, 95\%CI=30.6-36.2$, see Figure \ref{fig:harmsexperienced}). Overall, significant differences by gender were in the direction of greater harms to women, such as mental health harms ($ARR=1.69, 95\%CI=1.40-2.04, p<0.001$), experiencing abuse in other ways by the perpetrator ($ARR=2.0, 95\%CI=1.42-2.81, p<0.001$), and detrimental impacts to intimate relationships ($ARR=1.55, 95\%CI=1.14-2.11, p<0.01$). LGBTQ+ respondents were 1.69 times more likely to indicate their physical health had been affected as compared to non-LGBTQ+ respondents ($19.7\%$ vs $11.7\%, p<0.01$, not shown). 

Respondents who selected ``other'' to the question about harms were asked to provide more detail in an open-ended format. A large number of respondents (74.0\% of the 274 coded responses) reported no impact from their experience, using language such as ``indifferent'' and ``didn't care.'' For instance, a Dutch woman (aged 35-49) explained: ``He has kept all the footage to himself until now. It never came out. So I don't think it's a problem for his own collection that he still has it.'' Among those who indicated no impact, a common sentiment was a lack of concern, or a belief that their victimization had little to no effect on their daily life. Some mentioned that because their images were never distributed, they encountered no discernible harm: ``Nothing came out either on the internet or with my relatives, no one knows'' (Mexican woman, aged 18-24). Some respondents characterized their experience as trivial or harmless ``fun.'' For example, a man from Australia (aged 50-64) remarked, ``We’re all army boys who were having a laugh.'' These types of responses may have been less common had we included a ``no impact'' answer option.

In contrast, some respondents chose ``other'' to expand on harms they had experienced. The most commonly mentioned harm was mistrust in others. For instance, several respondents highlighted the ways that IBSA had impacted their relationships, noting issues of trust and feelings of betrayal. One woman from the USA (aged 18-24) explained how she felt uncertain and confused about her future relationships: ``[I have] more doubt and confusion about my [future] partners and how to tell whether they’re trustworthy or good people. [I also have] doubt in myself.'' One man from Mexico (aged 34-45) noted that ``I los[t] a relationship because it was implied that the photo was recent.'' The second most commonly mentioned harm was fear/anxiety, including concern about others finding out. For example, a Polish woman (aged 25-34) explained how she felt ``terrified that the material would be made public to a wider group of people.'' Finally, the third common theme was shame/embarrassment. For example, one woman from the USA (aged 35-49) mentioned how her victimization ``Exacerbated unhealthy behaviors - heavier drinking of alcohol to forget and suppress the shame/memory,'' while another respondent (USA woman, aged 25-34) said, ``I felt like I was played, and [was] disgusted with myself.''

 \begin{figure*}[!htpb]
	\centering
\includegraphics[scale=.6]{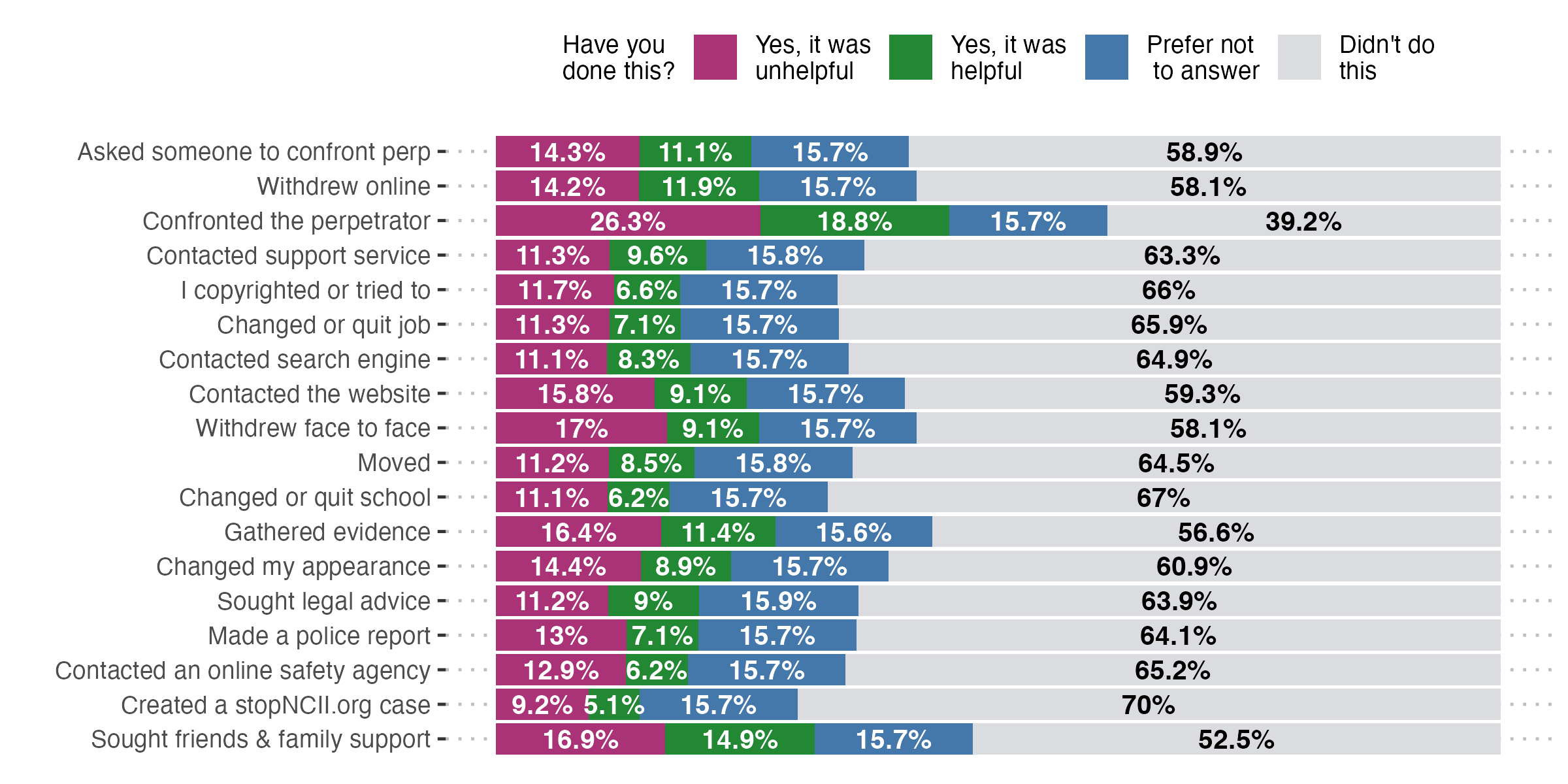}
	  \caption{Victim-reported actions taken, and whether they were helpful. Note that ``didn't do this'' includes both actions that the victim-survivor chose not to take, as well as cases where the action was not applicable.}\label{actionstaken}
\end{figure*}
\subsection{\textbf{RQ5: Help-Seeking and Formal Reporting Experiences}}

\begin{figure}[!htpb]
	\centering
\includegraphics[scale=.25]{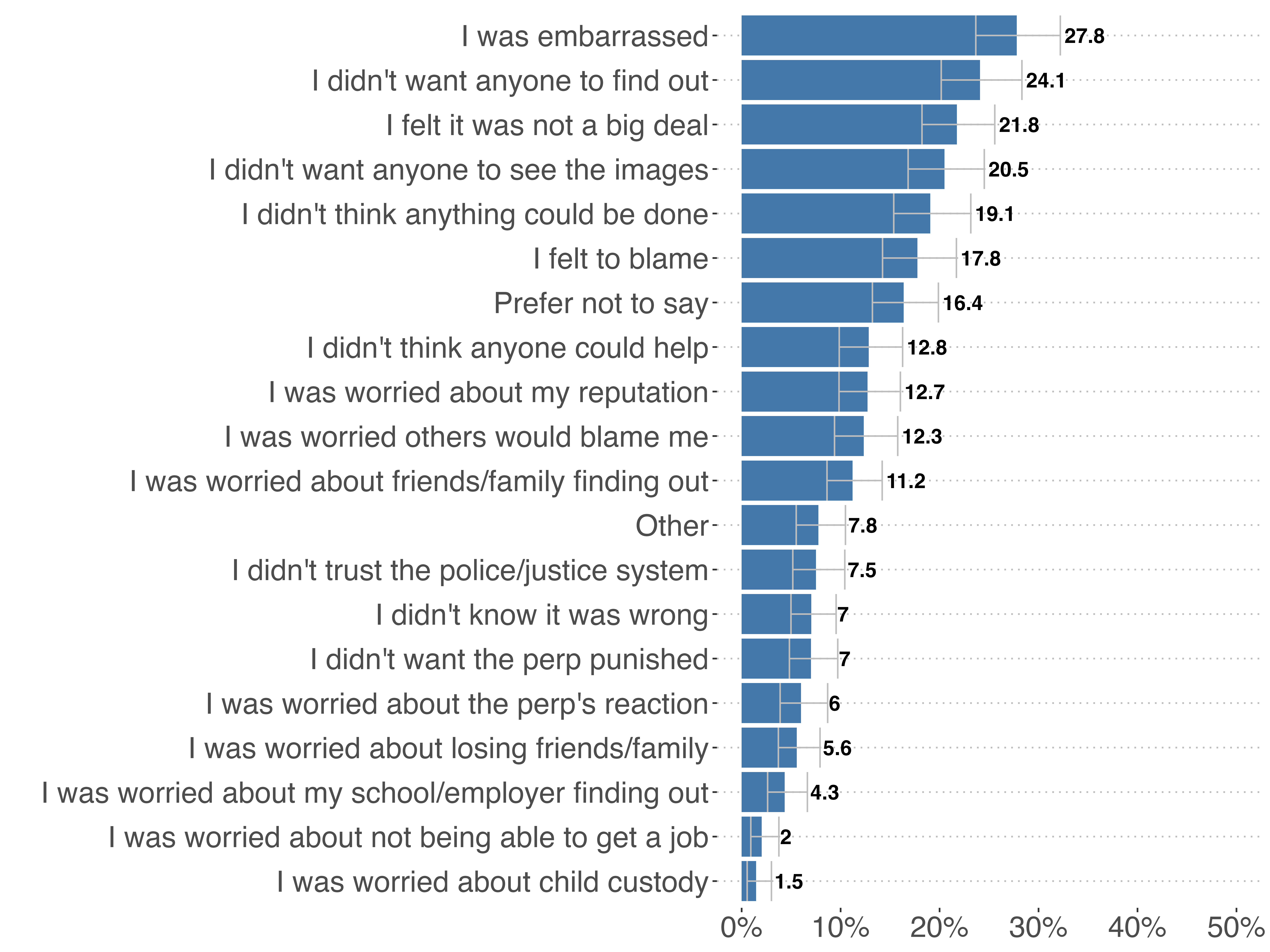}
	  \caption{Why didn't victim-survivors tell people/report their experience? (95\% CIs in grey).}\label{fig:nondisclosure}
\end{figure}
Respondents were asked to indicate which of a series of actions, if any, they had taken in relation to their IBSA experience. They were also able to skip as much of this question as they wanted. 
For all of the listed actions, respondents who took that action were more likely to report that it was unhelpful than helpful. 
Perhaps unsurprisingly, given the ubiquity of intimate/former intimate partner perpetrators, the most common action taken in response by victim-survivors was confronting the perpetrator, but again, most respondents found this to be unhelpful rather than helpful. 

To better understand the disclosure patterns, victim-survivors were grouped according to the stated actions they had or had not taken. In total,
38.8\% ($95\%CI=36.2-41.6$) of victims had reported to an agency, law enforcement, or digital platform (and may have also disclosed informally); 14.3\% ($95\%CI=12.5-16.3$) told someone but did not report; and 30.9\% ($95\%CI=28.5-33.4$) did not report or disclose to anyone. The remainder ($15.9\%, 95\%CI= 14.0-18.0$)
chose ``prefer not to say'' for all of the disclosure and reporting questions. 
 
\begin{table}[!htpb]
\sf\centering

\caption{Motivations for victim-survivors to report their experience}\label{tbl:reportingmotivations}
\begin{tabular}{lr}
\toprule
\textbf{Reason} & \textbf{\% of Reporting} \\& \textbf{Respondents (95\% CI)}\\
\midrule
Take down the imagery & 36.3\% (31.9-40.7) \\
Punishment of perpetrator & 32.5\% (28.2-37.0)\\
Perpetrator faces legal system & 27.3\% (23.4-31.5)\\
Emotional support & 27.0\% (23.0-31.3) \\
Identification of perpetrator & 24.3\% (20.5-28.4) \\
More information about options & 23.5\% (19.6-27.7)\\
Recognition of my experience & 20.1\% (16.6-24.0) \\
Prefer not to say & 9.2\% (6.6-12.4) \\
Other & 1.5\% (0.6-3.0) \\
\bottomrule
\end{tabular}\\[10pt]
\end{table}

Among the respondents who reported IBSA, more than half ($52.2\%, 95\%CI=47.6-56.8$) reported multiple motivations for not taking action (not shown, see Table \ref{tbl:reportingmotivations} for individual reasons). As seen in Figure \ref{fig:nondisclosure}, those who did not tell anyone or report their experience to authorities commonly endorsed reasons echoing sentiments seen in the open-ends. For instance, in the survey, the ``other'' option provided in the ``what actions did you take'' question was intended to allow respondents to indicate any actions they had taken beyond those listed. The most common ``action'' taken by the 439 respondents who chose the ``other'' category was no action at all (36.0\%). A common reason was that their victimization occurred some time ago when there were fewer available resources or avenues for addressing the problem. Similarly, respondents noted that because the perpetrator had not distributed the image, they felt they lacked the authority or grounds to take further action. For example, a woman from the Netherlands (aged 35-49) remarked, ``Nothing was ever posted online, so I couldn't do anything.'' Additionally, the fear of being shamed or perceived as overreacting contributed to a reluctance to pursue further action. A man from Mexico (aged 35-49) reflected on how his views on masculinity prevented him from seeking recourse: ``As a man ... I did not say anything or do anything so as not to be seen as weak.'' 

Some respondents described their preference to handle the situation independently. For instance, some felt confident in their ability to resolve the issue themselves, while others believed that managing it privately would reduce the risk of their victimization being discovered by others: ``I thought it was embarrassing and didn't want him - or even worse - others to see it'' (Danish woman, aged 50-64). 

Additionally, respondents used the ``other'' open-end as an opportunity to expand on existing listed actions. The most commonly mentioned action was contacting the perpetrator. A man from the USA (aged 25-34) explained how he ``personally met [with the perpetrator] and pleaded with her to delete [the images] from her device.'' Some respondents also confronted the perpetrator to demonstrate the extent of harm they had experienced: ``I tried to get him to understand how humiliating and abusive his actions were'' (Australian woman, aged 25-34). The second most commonly mentioned action was ceasing or limiting communication with the perpetrator and changing social circles to completely ``avoid contact'' with the perpetrator. And the third most commonly mentioned action was taking down or deleting content. For instance, a woman from the USA (aged 18-24) said, ``I deleted the videos on the WhatsApp chat and asked him to do the same. I researched it online to see if it could be taken from our phones. I told him not to show anyone.'' Taken together, these open-ended responses support the quantitative data, while providing deeper insights into respondents' actions and personal experiences.

\section{Discussion}
In this study, we investigated the prevalence, nature, and impacts of different forms of IBSA, including the nonconsensual taking, creating, and sharing of intimate images, and threatening to share intimate images. As earlier noted, this is the largest and most comprehensive multi-country survey on IBSA to date. 

\subsection{Prevalence: Cross-Country Comparisons}

Almost a quarter of respondents indicated at least one IBSA victimization experience, with the lowest rates seen in European countries. Just over an eighth of respondents indicated multiple types of IBSA victimization. The most common experience reported was being threatened (sextortion), and the least common was the creation of digitally altered or manipulated intimate images. These relatively high rates may still represent an underestimate, given potential lack of knowledge of victim-survivor status (e.g., when images are taken, created, or shared surreptitiously without the victim's knowledge), as well as intentional under-reporting. 

\color{black}

Comparisons to results from previous studies should be treated with caution for several reasons, including the likelihood of changing rates \citep{ThornNCMEC2024, ward2021intimate}, a lack of rigorous prevalence data in the majority of countries we surveyed, and \color{black}
 methodological differences between studies, including definitions, the level of granularity in reported data, and timeframes of data collection. More similar studies include those on nonconsensual sharing conducted by \citet{ruvalcaba2020nonconsensual} and the Australian \citet{esafety2017} who, like us, surveyed a national sample of adults in the United States and Australia, respectively. 
Our study found a higher rate of victimization for nonconsensual sharing among American respondents (14.0\%) compared to American respondents in \citet{ruvalcaba2020nonconsensual} (8.02\%). We also found a higher rate of IBSA among Australian respondents (15.9\%) in comparison to the study conducted by the \citet{esafety2017} (11.0\%). 
In addition, our study found equivalent rates of victimization for men and women, whereas the two aforementioned studies found more women than men experienced this. 
 The increased rates compared to prior studies may reflect both an actual rise in IBSA victimization and the impact of the COVID-19 pandemic, which drove  which drove increased digital technology and internet \citep{feldmann2021year}. In relation to the difference between men and women, the recent trend of men being targeted for sextortion \citep{o2022cyber, C3P, cbs} may explain the overall higher rates of IBSA among men in our study compared to \citet{ruvalcaba2020nonconsensual} and the study by the \citet{esafety2017}.

 Only one other multinational study on the prevalence of IBSA has been conducted to date \cite{henry2020image}. In that study, conducted prior to COVID-19 and with individuals aged 16-64, three relatively similar countries were surveyed (Australia, New Zealand, and the UK). Overall, the study found that prevalence of IBSA (across three behaviors) was slightly higher in New Zealand and the UK (39.0\%), compared to Australia (35.2\%). They also found, differently to our study, that the most common form of IBSA was the nonconsensual taking of intimate images, followed by nonconsensual sharing and then threats to share intimate images. By way of comparison, in our study, we surveyed a larger number of countries that were also from three separate continents, and our population had a more representative sample of adult age. We saw differences in rates between countries, and within demographic groups, indicating that cultural and societal factors are likely to play a role. A general trend was that European countries and South Korea had the lowest rates of victimization, while Mexico, Australia, and the United States had the highest rates. Below we touch on several potential reasons for the geographical differences, with the caveat that exploring every possibility is beyond the scope of this study. 
\subsubsection{Societal Views on Sexuality and Nudity}
One possible explanation for differential rates, and in particular generally lower rates in the European countries surveyed, may be due to more permissive societal attitudes about sexuality and nudity in many European countries \citep{fortenberry2022sexual}, such that nude or sexual images are less easily weaponized for stigmatization and shaming. 

\subsubsection{Relationships and Sexuality Education}
Relationships and sexuality education may also play a role, particularly if concepts like care, empathy, consent, and respect are not taught or are introduced late \citep{burton2023teaching}. While there is some variability, European countries generally have some form of mandatory relationships and sexuality education, often starting young, with little emphasis on abstinence \citep{parker2009sexuality}. In Australia, relationships and sexuality education are taught within schools from preschool to Year 12, 
including age-appropriate consent education from the first year of schooling to Year 10. Sexting education, however, continues to focus on abstinence in Australia \citep{woodley2024send}. In the United States and Mexico, sexuality education is a controversial topic that has become increasingly polarized \citep{chandra2018evolution, hall2016state}. Even within countries, the curriculum may vary significantly. This is evident in the United States, where in many states sexual education is not mandatory and ``abstinence only until marriage'' remains the default approach \citep{caulfield2024considering}.

\subsubsection{Gender Inequality}
We might expect gender inequality would influence victimization across the different countries. However, ranking on the World Economic Forum's 2020 Global Gender Gap Index \citep{wef_global_2023} or on the UN Development Programme's Gender Social Norms Index \citep{hdr2023} was not predictive of victimization rates in our study. As has been well-documented in the intimate partner violence literature, the relationship between gender equality and gender-based violence is complex \citep{gracia2016intimate}.

\subsubsection{Laws}
Most of the 10 countries we surveyed have introduced laws to criminalize the nonconsensual sharing of intimate images, but not all countries have specific criminal laws on other forms of IBSA, including the nonconsensual creation of digitally altered or manipulated intimate images (``deepfakes''), or threats to share intimate images (``sextortion''), although there may be broader laws that can be used (e.g., blackmail). Also, within countries that have a federal system of governance, including Australia, Mexico, and the United States, there are variations and inconsistencies between jurisdictions. For instance, with the exception of Tasmania, all Australian state, territory, and federal jurisdictions have specific IBSA legislation. Mexico similarly has criminal offenses at the state and federal levels, but not all states have introduced laws. In the United States, 48 states, Washington D.C., and two territories have IBSA laws (and, at the time of writing, the ``SHIELD'' Act had been passed in the Senate to make the distribution of private intimate images a federal crime). 
 
The role of law in deterring and preventing crime from occurring is contested. As noted by \citet[p.~70]{larcombe2014limits}, law reform has had little impact on the prevalence of sexual assault owing to the lack of an effective threat of punishment because of low prosecution and conviction rates, as well as the ``conflicting attitudes, values, and norms in social spheres beyond the legal sphere'' that may be resistant to the legal norms that new laws seek to establish. This is also the case with child sexual abuse laws that criminalize minors sexting consensually. Given the prevalence of sexting among youth \citep{Power2022}, these laws seem relatively ineffectual in preventing youth sexting. Similarly, given the high rates of IBSA in some countries with comprehensive IBSA legislation and/or significant penalties under criminal and/or civil regimes, the laws may not have the effect of reducing IBSA prevalence, especially if they are not adequately applied in practice or if police and other authorities have little knowledge of such laws. 
\subsection{Risk Factors for Victimization}

\subsubsection{Gender}

IBSA is often considered a form of technology-facilitated gender-based violence because of the disproportionate impacts on women. Gender, however, plays a complicated role in both the likelihood of experiencing IBSA, and the experience itself. In our study, the aggregated rates of victimization were equivalent for men and women, as were rates for nonconsensual recording and distribution. Men reported higher rates of certain subtypes of IBSA consistent with higher rates of sextortion victimization reported by men in several other studies \citep{eaton2023relationship, gamez2023technology, henry2019image, henry2020image, patchin2020sextortion}, although it differs from a recent study of adolescents, which found equivalent rates between boys and girls \citep{patchin2024nature}. 
Gender differences within countries did surface in both directions (e.g., South Korea, Australia, Belgium, France). These disparities highlight the diverse motivations of perpetrators and subtypes of IBSA, and the importance of cultural factors.

Given the importance of such results in terms of funding and resourcing, we take care to note two complications in these results. The first is that women may be less likely than men to become aware of their victimhood status; for instance, when intimate images are ``privately'' shared by perpetrators for peer-to-peer bonding or to demonstrate sexual prowess. And second, as noted in section 5.4 below, victim gender influences perceptions of harm and impact severity.

\subsubsection{Sexuality}
Echoing previous findings \citep{ruvalcaba2020nonconsensual,brighi2023prevalence, henry2019image, karasavva2022personality, esafety2017}, in our study we found that LGBTQ+ respondents were almost two times more likely than non-LGBTQ+ respondents to report IBSA victimization. An explanation for this may be systemic discrimination on the grounds of gender and/or sexuality, higher use and engagement in online dating and social media, and intimate images being used as a means to ``out'' or expose LGBTQ+ status \citep{mitchell2023navigating, powell2020digital}. In our study, we did not find an interaction between gender and LGBTQ+ status as LGBTQ+ women reported similar rates as compared to LGBTQ+ men. This is in contrast to \citet{ruvalcaba2020nonconsensual}, who found that bisexual women were more likely than any other group to have experienced IBSA, including bisexual and gay men (although it is worth noting that we did not differentiate between gay/lesbian and bisexuality for our LGBTQ+ analyses). 

\subsubsection{Age}
Young people were much more likely than older respondents to report victimization. This is consistent with existing studies demonstrating that emerging adulthood is a significant risk factor for IBSA \citep{eaton2023relationship, ruvalcaba2020nonconsensual, henry2020image, esafety2017}. Youth represent a convergence of risk factors: an adolescent brain with underdeveloped executive functioning \citep{best2010developmental}, the onset of sexual and romantic relationships, early access to and adoption of emerging technologies \citep{hauk2018ready} such as smartphones and social media \citep{ehman2019sexual}, and limited forms of currency to build relationships and establish their place in the social structure. Our findings underscore the importance of comprehensive relationships and sexuality education, as well as specific programming on the topic of image sharing. 
\subsection{Relationships}
Consistent with the literature, we found perpetrators were most often former or current intimate partners \citep{branch2017revenge, ruvalcaba2020nonconsensual, henry2020image, dardis2022nonconsensual, henry2024sextortion, esafety2017}. The exchange of intimate images (whether consensual or coerced) is common, particularly amongst youth and emerging adults \citep{ Power2022}. This behavior is considered developmentally normative and many experts agree that rather than promoting abstinence, educational strategies should work primarily to educate would-be perpetrators and embrace harm reduction messaging and digital safety strategies \citep{qin2024didfingconsentthat, qiwei2024feministinteractiontechniquesdeterring}. 

\subsection{Harms and Actions Taken}

Victim-survivors most commonly reported deleterious effects to their mental health as a result of experiencing IBSA, but negative impacts were relatively wide-ranging, touching their personal and professional lives. 
However, for some people, there were little to no impacts, or they found the experience to be flattering, sexy, or funny, a finding also observed by \citet{henry2020image}. \color{black} 

Although aggregated prevalence across genders was equivalent, gender nonetheless played an important role in the feelings and harms associated with experiencing IBSA. The findings therefore support gender as a moderator of the consequences of nonconsensually shared images due to cultural expectations in gendered societies \citep{ringrose2022wanna, rackley2021seeking} and echo previous findings \citep{dunn2020technology, henry2020image}. While banter and humor reduces the social stigmatization of leaked ``dick pics,'' girls are subjected to ``slut shaming'' and social stigmatization when their intimate images are shared. The context of the victimization may also play an important role here--for instance, the sharing of intimate images amongst heterosexual men as a form of homosocial bonding has been documented by several scholars \citep{hall2019revenge, henry2019image}. 

In relation to reporting and help-seeking, previous research has been very limited. Most survey studies investigating formal reporting to authorities or digital platforms have centered on children and adolescents \citep{patchin2020sextortion, wolak2018sextortion}, or have focused on technology-facilitated violence more broadly \citep{Flynn28052023}, or specific types of IBSA, such as the nonconsensual sharing of intimate images \cite{ruvalcaba2020nonconsensual}. Concerning help-seeking, one study found that most victim-survivors ($72.95\%$) did not ``turn to anyone for help'' \citep{ruvalcaba2020nonconsensual}; another found that nearly half of the victim-survivors they surveyed ($43.9\%$) ``did not tell anyone'' about their experiences \citep{brighi2023prevalence}; and another study found that two-thirds had ``low confidence knowing where to go for help'' \citep{henry2020image}. 

In our study, we included several questions on reporting and help-seeking to understand victim-survivor behaviors and barriers. For instance, we asked about both reporting formally and disclosing informally, finding that almost a third of victim-survivors did not disclose or report their experience ($30.9\%$). We also found that a majority of victim-survivor respondents did not formally report to an agency, law enforcement, or a digital platform ($61.2\%$).

While studies have examined help-seeking on social media \citep{wei2024understandinghelpseekinghelpgivingsocial} or in relation to online abuse more broadly \citep{dunn2023supporting}, our study was the first to comprehensively investigate help-seeking behaviors among IBSA victim-survivors. Our study contributes useful insights into the specific actions taken that other studies have not yet investigated, including telling a trusted person, reporting to authorities, and confronting the perpetrator, among others. For instance, across the actions listed in our survey, the majority of respondents indicated they either had not done it, or it was unhelpful when they did do it.

As our study shows, the failure to take any action in relation to IBSA may stem from shame, a lack of awareness, deficient recourse options where they live (particularly for older experiences that may have occurred before relevant laws were passed), or an unwillingness or inability to deal with many of the structural barriers associated with taking action. Respondents also indicated that they felt that taking action would not have actually been helpful. \color{black} The most common action taken in response to victimization was confronting the perpetrator, which when done, was considered helpful approximately half of the time. The comparative frequency of this action may reflect both that the perpetrator was often a current or former partner, and that victim-survivors may want perpetrators to recognize their wrongdoing and take accountability \citep{rackley2021seeking}. This finding has important implications for the design of help-seeking resources, tools, and interventions.

\color{black}

The overall results regarding the efficacy of help-seeking and reporting are sobering. Even those respondents who took action were less likely to report it being helpful than unhelpful, indicating that the options available to victim-survivors fall short. While primary prevention remains a priority, more research is needed on remediation measures.

\subsection{Design Considerations}

The findings of this study have important implications for HCI research, policy, and practice. 

\subsubsection{Research}
Recognizing the relatively limited data available regarding prevalence, this research set out to help fill that gap by providing datapoints in multiple countries for different subtypes of IBSA. The necessity of more granular measurement of a wider set of behaviors becomes clear when we consider differences that only emerged when examining the data by subtype of IBSA (e.g., equivalent rates of IBSA victimization among men and women, but higher rates of images being created or stolen for men). \color{black} While we note country differences, future research conducted by researchers with cultural expertise will be better able to test explanatory factors. 

The relative lack of data on IBSA calls for better and more comprehensive measurement. Existing community-based crime victimization surveys, such as the annual American National Crime Victimization Survey \citep{ncvs} and the Australian Bureau of Statistics \textit{Personal Safety Survey}, could create or expand on IBSA-dedicated questions to facilitate longitudinal trend tracking. The internet is not necessary to perpetrate IBSA, but advances in digital technologies have facilitated IBSA at every stage, from easier capturing and creating of images, to effortless, widespread transmission of such images. Longitudinal tracking can help: a) identify the impacts of changing legislation and affordance/application development; and b) surface trends. We underline the importance of language in crafting these questions. According to our results, many individuals may not necessarily think of themselves as victims if they did not experience negative effects.   

Although survey research is useful for quantifying patterns and trends, more HCI qualitative research with diverse participants is also needed to provide deeper understanding of the impacts of IBSA victimization, as well as the barriers and facilitators to help-seeking and reporting (including how intersectionality shapes not only abuse experiences but also help-seeking and reporting). There is a significant lack of qualitative research with men respondents, which could probe some of the trends around impacts and harms seen in our quantitative data. Additionally, our efforts to provide baseline data in new countries should be extended through targeted funding of native researchers with cultural expertise.

More research is also needed to understand the motivations, attitudes, and behaviors of perpetrators. Research on victimization and perpetration can help inform much-needed developments in policies and practices to take the burden off victim-survivors, shape appropriate accountability measures, and to prevent IBSA perpetration.
 Crucially, these recommendations for research agendas should be considered as part of an overall call for greater cross- and inter-disciplinary collaboration in the field of IBSA, especially as evolving technologies facilitate scaled and online harms (e.g., financial sextortion by organized crime syndicates, ``deepfake pornography,'' and largescale groups dedicated to the nonconsensual dissemination of intimate images). Within the fields of criminology, sociology, gender studies, and legal studies, scholars have attempted to estimate prevalence, measure harms, and consider ways to mitigate harms through addressing the criminal justice system \cite{mcglynn2017beyond, mcglynn2017image}. Computer science and HCI researchers have also studied IBSA, but often via a privacy and security perspective \cite{sambasivan2019they, qin2024didfingconsentthat}, and focusing on ways in which technology may serve a helpful purpose in help-seeking \cite{wei2024understandinghelpseekinghelpgivingsocial, andalibi2018social}. Interdisciplinary collaboration strengthens research and leads to more comprehensive, context-informed recommendations. As an example, prevalence data provides more information about the full extent and the potential harm of IBSA, a crucial consideration for recommendations to technology companies and legislators. \color{black}

\subsubsection{Policy and Practice}
Relatively high rates of victimization in Australia and the United States, in particular, complicate the narrative that IBSA will be less prevalent in places with higher rates of gender equality (and thus complicate the suggestion that reducing gender inequality will result in reduced IBSA). A more targeted recommendation is the development or continuation of comprehensive relationships and sexuality education. This should include educational interventions focused on digital and porn literacy, harm reduction, affirmative consent, dismantling problematic gendered norms, attitudes, and beliefs, and recognizing the harms of IBSA and gender inequality more broadly. According to \citet{waling2024dude}, young people understand consent but they do not always practice it in the moment. They argue that in addition to the curriculum engaging with pleasure and positive sexuality, ``a broader, whole-of-society approach to thinking about sexual communication and sexual consent is needed'' \citep[p.~70]{waling2024dude}. This includes shifting attitudes underpinning masculine entitlement to women’s bodies, the denigration of women and gender-diverse people, the shame and stigma attached to their sexuality, and empowering all genders to celebrate and enjoy their sexuality. Moving from abstinence models to harm reduction models  will help people engage in sexting and other behaviors in safer and more respectful ways \citep{woodley2024send}. While sexuality and respectful relationships education is usually considered in the context of a school-based curriculum, there is a growing body of online resources. Currently, NGOs \citep{fumble, amaze, ccri2024safetycenter} and technology companies \citep{facebook2017notwithoutmyconsent, facebook2024instagramsextortion, bumble2024thebuzz} create and surface educational and wellbeing resources directly on the web for greater reach (including translated content).
Considering the spectrum of information quality in this space, reputable resources must be given place of primacy \citep{granka2004eye}.  

\color{black} 

 Laws are also an important part of the solution to IBSA, however our findings indicate that IBSA is relatively high even in those countries with strong laws (e.g., Australia, South Korea). This apparent poor deterrent effect is consistent with criminological literature that observes poor general deterrence when potential offenders may be both ignorant of laws and/or unconvinced they will be sanctioned under them \citep{robinson2004does}. Given the finding that perpetrators are often former or current intimate partners, the low rates of reporting to the police, and that victim-survivors most commonly confront perpetrators, it is important that alternative justice pathways are in place to respond to IBSA, tailored accordingly to cater to the specific needs of minoritized groups. Considering under-reporting and nondisclosure trends for IBSA and sexual violence generally, digital interventions (anonymous, private, self-serve) are an essential part of redress measures \citep{gorissen2023online}.

\color{black}

\subsubsection{Digital Interventions}
The high prevalence of victimization found in our research underscores the need for technology companies to contribute to solutions,
many which have been detailed previously \citep{qin2024didfingconsentthat, qiwei2024feministinteractiontechniquesdeterring, hamilton2023safer, saferdigital, qiwei2024sociotechnical}. Other HCI scholars note the importance of designing more inclusive, user-centered, and safer technologies to prevent technology-facilitated abuse happening in the first place \citep{mckay2021standing, freed2018stalker, brown2024safeguarding}. A ``safety by design'' stance might require inclusion of threat modeling and red-teaming in the technology development process, to ensure protective elements and affordances are built-in and intentional. In addition to these preventative considerations, ensuring that remediation tools are human-centered and trauma-informed can help victim-survivors take advantage of existing and new tools for redress \citep{chen}.
Crucially, there needs to be widespread adoption of affordances by digital platforms, especially as people tend to sext on the platforms they are already using \citep{coduto2024delete, hamilton2023safer}. Additionally, given the observation that desire for privacy may be misconstrued in the context of a relationship or in specific cultures \citep{umbach2023your, walton2012degrees}, and that perpetrators are most commonly current or former intimate partners, a privacy-forward stance with regard to feature defaults (e.g., opt-out rather than opt-in) will normalize the use of privacy-forward affordances. Below we focus on a range of affordances and digital tools that have been or could be introduced. We segment recommendations into the subtype of abuse they may be most useful in preventing, noting that some affordances fall into multiple categories.

\paragraph{Creation of Intimate Images} Many companies have already introduced features to prevent the nonconsensual creation of intimate images, although adoption levels vary. Affordances in this category include: notifications or blocking of screenshotting and recording (to eliminate permanent capture of content intended to be ephemeral) and forced shutter sounds on devices to deter ``creepshots'' \citep{tokyoweekender2024shuttersound}.

\paragraph{Threatening to Share Intimate Images}
Recognizing that some online harms involve private messaging contact between strangers, many platforms  provide useful information to the message recipient about the sender. For example, indicating whether a message sender is an existing contact, whether you have additional connections (e.g., participation in the same group), and the country of the sender \citep{whatsapp}. In the context of financial extortion, typically by a stranger, target hardening can include reducing visibility of social connections (e.g., follower lists), and detection and flagging of potential grooming patterns \citep{instagram}. Some platforms have also introduced on-device detection and flagging of nude images \textit{before} sending, to prompt users to pause and consider potential pitfalls before sharing \cite{googlenude, saferdigital}.

\paragraph{Sharing of Intimate Images}
Reducing the nonconsensual sharing of intimate images should be considered from several angles, and have corresponding recommendations. Here, rather than discussing standard security recommendations, we focus on more targeted recommendations. Unauthorized access to images via shared devices can be reduced through the use of ``locked'' or ``hidden'' folders within photo applications, which add an additional layer of security \citep{geeng2024say}. ``Shoulder surfing,'' where photos may be exposed accidentally by the recipient due to sharing or unattended devices, can be reduced via 
automated detection and masking of explicit images (requiring the recipient to take an additional step to view the content) \citep{googlenude, discord}. Applications with ephemeral messaging capabilities can also be used to reduce long-term retention of images. An additional and relatively new tool is stopNCII.org, which allows individuals to upload and digitally hash images, which can then be used to prevent upload to various platforms \citep{stopncii}. Like an ``unsend'' feature, this recognizes that users may want to withdraw access to images once shared consensually \citep{coduto2024delete}. Sites that specialize in explicit material (e.g., PornHub, OnlyFans) leverage facial recognition technology to prevent upload of explicit images by individuals not in the content \citep{onlyfans, mashable}. 
In addition to these interventions targeted directly at the involved parties, interventions can empower bystanders by educating and equipping them with the language and tools to intervene. More technically, platforms might consider the feasibility of expanding reporting permissions beyond victim-survivors and their authorized representatives, though we acknowledge that bad actors abusing such a policy might render this impractical. 
\paragraph{Help-Seeking, Resources, and Secondary Prevention}
The high prevalence of IBSA victimization found by our study and its potential to result in significant harms underscores the importance of providing victim-survivors with confidential, accessible, and trauma-informed information, support, and reporting options \citep{glass2017longitudinal, gorissen2023online}, which should be a priority for policymakers and practitioners \citep{chen}. Our survey results suggest that existing tools are relatively under-utilized or unhelpful, thus more research is needed to understand how to redesign, better surface, and add to existing methods of redress. We found that feelings of embarrassment and helplessness are drivers for non-disclosure and non-reporting, which suggests the ongoing need for anonymous, alternative reporting and support options, such as chatbots \citep{falduti2022use, maeng2022designing, henry2024design}. Additionally, mutual help platforms have already proved helpful in a range of online harms, including IBSA broadly \citep{wei2024understandinghelpseekinghelpgivingsocial}, as well as specific types of IBSA \citep{C3P}. While stopNCII.org helps prevent widespread distribution of nonconsensual images, carefully designed tools leveraging facial recognition might be explored as a way for victim-survivors to find and report more comprehensively across the web (e.g., images of which they were unaware). Our findings around gender prevalence in subtypes of IBSA and harms, specifically, suggest that help is not always perceived as needed, and equity must be considered in prioritizations. We reiterate the call by previous HCI researchers for a trauma-informed approach \citep{chen} in developing these interventions, and in particular, digital resources.

\subsection{Limitations}
This research offers valuable insights into cross-cultural trends on IBSA in 10 different countries, however, we note limitations and challenges. While we intentionally sampled a diverse range of countries, more localized investigations by researchers with the requisite cultural knowledge and expertise is needed to understand the rates and impacts of IBSA victimization in those countries. Similarly, while our study is the largest of its kind to date, our country choices are still disproportionately WEIRD (Western, Educated, Industrialized, Rich, and Democratic) due to ethical and methodological concerns. Indeed, other surveys on the topic of technology-facilitated gender-based violence have explicitly refrained from asking questions about intimate images in certain countries due to similar concerns \citep{dunn2020technology}. Nevertheless, we note that a more globally representative sample may report different findings, particularly around help-seeking (where certain options may be less relevant), or if definitions of intimate are more expansive. Local researchers would be best situated to conduct this type of research safely and ethically.

In addition to the likelihood of undercounting of unaware victim-survivors, sensitive topics can lead to intentional under-reporting and selection bias. Several aspects of our study were intended to minimize this risk. For example, anonymous surveys have less under-reporting bias relative to qualitative interviews \citep{cullen}. Respondents could choose ``prefer not to answer'' for any of the more sensitive questions, and still complete the survey. Questions were worded to be descriptive and valence-free, to ensure we captured instances that were not perceived as harmful or abusive. Nevertheless, we recognize that cultural norms around sexuality and taboos, as well as general awareness of IBSA, may influence reporting. This is true of participating in the survey, reporting IBSA in general, and reporting harms/impacts experienced. The respondent sexuality results in South Korea may indicate our attempts at mitigating non-response bias were less successful in this traditionally conservative country \citep{cha2023time}, and may help explain the observed gender difference. Future research looking to replicate our findings or probe their robustness could include IBSA questions as part of a larger study on online harms.

Due to the length of our survey, we used a single matrix-style question to ascertain IBSA subtype victimization status. It is typically preferable to ask additional questions in case victim-survivors change their mind about reporting, or remember something during the process. However, unlike many other studies, we asked individual questions about the different types of victimization, which should lessen this concern particularly when it comes to the aggregated rates of victimization. The length of our study also precluded our ability to ask about additional forms of gender-based violence or technology-facilitated violence, which often co-occurs \citep{backe2018networked, dunn2020technology}. 

While we did find heightened rates of victimization for LGBTQ+ respondents in line with previous literature, men and women within that group reported equivalent rates. Future research could consider oversampling of at-risk minoritized populations to plumb the relationship between gender and sexuality. 

Finally, while we had hoped that answer options for follow-up questions (e.g., How did you feel about this experience? What impacts did it have?) were exhaustive, respondents most commonly used the ``other'' category to indicate that none of the above were applicable, because they had no feelings or there were no harms. Nevertheless, the elaborations on the lack of perceived harm provided useful insights. Future studies may consider inclusion of a ``none'' category, paired with an open-ended field that allows respondents to expand further.

\section{Conclusion}

Our study on IBSA surveyed over 16,000 respondents across 10 different countries. This is the largest representative sample of adults on the topic. Prevalence rates for at least one experience of IBSA varied from a low of 13.6\% to a high of 29.9\%, averaging out at 22.6\%, or almost 1 in 4 adults. Our results echo some of the more consistent findings in the extant literature (e.g., LGBTQ+ and young respondents report disproportionately higher rates of victimization), and identify a potential cause for mixed findings regarding gender, namely the ways in which IBSA is measured and defined. Notably, while men and women report similar overall rates of victimization, men are more likely to report certain victimization experiences, such as the creation of imagery and being threatened with dissemination. This highlights the importance of measuring different forms of IBSA. Moreover, our study also found, consistent with other studies, that women report disproportionate harms as compared to men. 

 Previous research suggests that online tools (anonymous, free, immediately available) may be of particular help to victim-survivors, however our findings indicate that help-seeking by victim-survivors, including through online mechanisms, is still somewhat limited. More HCI qualitative research, particularly with at-risk populations, should be conducted for better understanding of harms and impacts, as well as trauma-informed methods to promote reporting and help-seeking \cite{chen}. \color{black}

\clearpage
\bibliographystyle{ACM-Reference-Format}
\bibliography{sample-base}
\input{demographics}

\end{document}

%% file: prevalenceoverall.tex
\begin{tabularx}
{.8\textwidth}
{lrrrr}
\toprule
 \textbf{Country} & \multicolumn{3}{c}{\textbf{Percentage of respondents (95\%CI)}} & \textbf{Between country differences}   \\
&  \textbf{Women} &  \textbf{Men{*}} & \textbf{Total} & \textbf{in total prevalence{*}} \\
\midrule
Australia$_a$ & 22.5\% (19.8-25.4) &   26.6\% (23.2-30.2) & 24.5\% (22.3-26.7) & \underline{b}, \underline{c}, \underline{d}, \textbf{e}, \underline{f}, \underline{g}, \textit{h}, \textit{i}\\
Belgium $_b$    & 13.4\% (11.2-15.7)& \textbf{18.4\%} (15.6-21.4) &   15.9\% (14.2-17.8) & \underline{a}, \underline{e}, \underline{j}  \\
Denmark $_c$   &   16.2\% (13.8-18.8)&  18.7\% (16.1-21.5) &    17.6\% (15.8-19.5)& \underline{a}, \underline{e}, \underline{j}\\
France $_d$     &  15.1\% (12.7-17.6)&  18.2\% (15.6-21.1) &  16.7\% (14.9-18.6)  &\underline{a}, \underline{e}, \underline{j} \\
Mexico $_e$ & 29.4\% (26.5-32.5) & 29.9\% (26.7-33.1) &    29.8\% (27.7-32.0) & \textbf{a}, \underline{b}, \underline{c}, \underline{d}, \underline{f}, \underline{g}, \underline{h}, \underline{i}, \textbf{j}  \\
Netherlands $_f$& 13.5\% (11.4-15.9)  & 15.9\% (13.4-18.6) &   15.0\% (13.4-16.8) & \underline{a}, \underline{e}, \underline{j}  \\
Poland $_g$          & 12.5\% (10.4-14.7) & 14.6\% (12.2-17.3) &    13.6\% (12.0-15.3)   & \underline{a}, \underline{e}, \textit{h}, \textit{i}, \underline{j}\\
South Korea $_h$&  24.0\% (21.0-27.2)    &\underline{13.6\%} (11.5-16.1)&18.8\% (16.9-20.8) & \textit{a}, \underline{e}, \textit{g}, \textit{j}\\
Spain$_i$ &18.7\% (16.3-21.4)&18.6\% (16.0-21.5)  &18.7\% (16.9-20.6) & \textit{a}, \underline{e}, \textit{g}, \underline{j}\\
USA$_j$&23.8\% (21.3-26.5)&24.8\% (21.5-28.3)&24.2\% (22.2-26.3) & \underline{b}, \underline{c}, \underline{d}, \textbf{e}, \underline{f}, \underline{g}, \textit{h}, \underline{i}\\
\bottomrule
\end{tabularx}

%% file: genderxsubtypes.tex
\begin{table}[!htbp]
\sf\centering
\caption{Victimization of subtypes of IBSA by gender, adjusted risk ratio (ARR) compares women's likelihood of victimization compared to men within subtype. Significance levels (no adjustments applied) indicated by asterisks: \textbf{$*p<0.05, **p<0.01, ***p<0.001$ }}\label{tbl3}
\begin{tabular}{llrr}
\toprule
\textbf{Type of IBSA} &\textbf{Gender} & \textbf{\% of Respondents} &\textbf{ARR}\\ &&\textbf{(95\%CI)}\\
\midrule
Created & Women & 6.6\% & 0.7***\\&&(5.8-7.4) & (0.59-0.84)\\
& Men & 9.4\% & \\&&(8.3-10.6) & \\
Photographed & Women & 13.9\% & 0.97\\ /filmed&&(12.8-15.1) & (0.85-1.1)\\
 & Men & 14.5\% & \\&&(13.1-15.9) & \\

 Stolen & Women & 8.2\% & 0.73***\\&&(7.2-9.2)&(0.62-0.86)\\
 & Men & 11.3\%\\&&(10.0-12.6)\\
 Threatened & Women & 14.5\% & 0.86*\\&&(13.6-15.4)&(0.76-0.97)\\
 & Men & 15.7\%\\&&(14.3-17.2)\\
 Sent/shown & Women & 12.2\%& 0.99\\&&(11.1-13.3) &(0.86-1.14)\\
 & Men & 12.4\%\\&&(11.1-13.8)\\

\bottomrule
\end{tabular}\\[10pt]
\end{table}

%% file: demographics.tex
\setlength{\tabcolsep}{2.5pt}
\begin{table*} [!htb]
\sf\centering
  \renewcommand\thetable{A1} 
  \captionof{table}{Participant Demographics. (Gender may not sum to 100\% due to the small fraction of respondents who preferred to self-describe, not disclose, or reported as non-binary.)}
\centering
\begin{tabular}{lccccc}
  \toprule
  \textit{Country} & \textbf{Australia}&\textbf{Belgium}&\textbf{Denmark}& \textbf{France}&\textbf{Mexico}\\ 
  \hline
\textit{Language(s) survey was offered in} & English & French & Danish & French & Spanish\\
\textit&&Dutch&&\\
   \hline
  \textit{Respondent count} & 1651 & 1617&1639&1636&1714\\
  \hline
  \textit{Gender}& 51\% women & 50\% women & 51\% women & 52\% women & 51\% women\\
  & 48\% men & 48\% men & 49\% men & 46\% men & 48\% men\\
  \hline
  \textit{Respondent} \hspace{2mm} 18-24&10\% & 9\% & 9\% & 9\% &16\%\\
\textit{Age} \hspace{12mm}25-34 & 10\% & 9\% & 12\% & 8\% & 13\%\\
\hspace{17mm}35-44 & 11\% & 10\% & 5\% & 9\% & 12\%\\
\hspace{17mm}45-54 & 9\% & 10\% & 10\% & 10\% & 10\%\\
\hspace{17mm}55-64 & 9\% & 10\% & 8\% & 11\% & 8\%\\
\hspace{17mm}65+ & 12\% & 11\% & 15\% & 12\% & 4\%\\
\hline
\textit{Sexual}  \hspace{9mm}Heterosexual & 87\% & 85\% & 91\% & 85\% & 84\% \\
\textit{orientation} \hspace{4mm} Gay\slash Lesbian & 4\% & 4\% & 3\% & 4\% & 3\%\\
\hspace{17mm} Bisexual & 5\% & 6\% & 3\% & 4\% & 5\%\\
\hspace{17mm} Prefer to self-describe & 2\% & 2\% &1\% &2\% &2\%\\
\hspace{17mm} Prefer not to say & 3\% & 3\% & 3\% &5\% &6\%\\
\toprule
 \textit{Country} & \textbf{Netherlands}&\textbf{Poland}&\textbf{South Korea}& \textbf{Spain}&\textbf{USA}\\ 
  \hline
\textit{Languages} & Dutch & Polish & South Korean & Spanish & English\\
   \hline
  \textit{Respondent count} & 1665 & 1639&1641&1711&1780\\
  \hline
  \textit{Gender} & 51\% women & 52\% women & 50\% women & 52\% women & 51\% women\\
  & 49\% men & 47\% men & 50\% men & 48\% men & 47\% men\\
  \hline
    \textit{Respondent} \hspace{2mm} 18-24&10\% & 9\% & 9\% & 8\% &11\%\\
\textit{Age} \hspace{12mm}25-34 & 9\% & 10\% & 9\% & 8\% & 12\%\\
\hspace{17mm}35-44 & 9\% & 11\% & 11\% & 12\% & 10\%\\
\hspace{17mm}45-54 & 11\% & 9\% & 11\% & 12\% & 10\%\\
\hspace{17mm}55-64 & 9\% & 11\% & 13\% & 13\% & 10\%\\
\hspace{17mm}65+ & 14\% & 9\% & 5\% & 8\% & 11\%\\
\hline
\textit{Sexual}  \hspace{9mm}Heterosexual & 86\% & 86\% & 68\% & 89\% & 82\% \\
\textit{orientation} \hspace{4mm}Gay\slash Lesbian & 4\% & 2\% & 3\% & 3\% & 7\%\\
\hspace{17mm}Bisexual & 5\% & 3\% & 15\% & 4\% & 7\%\\
\hspace{17mm}Prefer to self-describe & 1\% & 2\% & 5\% & 1\% & 2\%\\
\hspace{17mm}Prefer not to say & 3\% & 7\%&10\% &3\% &4\%\\
\hline
\hline
\end{tabular}
\end{table*}